\documentclass[runningheads]{llncs}
\usepackage{graphicx}
\usepackage{subcaption}
\usepackage{caption}
\usepackage{float}
\usepackage{hyperref}
\usepackage{babel}
\usepackage[utf8x]{inputenc}
\usepackage[numbers]{natbib}
\usepackage{epstopdf}
\usepackage{color}
\usepackage{xcolor}
\usepackage{nag}
\usepackage{tabularx}
\usepackage{multirow}
\usepackage{pgfgantt}
\usepackage{tikz}
\usetikzlibrary{positioning, arrows.meta}
\usepackage[title]{appendix}
\usepackage{booktabs}

\begin{document}
\title{Enhancing Medical Learning and Reasoning Systems: A Boxology-Based Comparative Analysis of Design Patterns}
%
%
\author{Chi Him Ng}
\authorrunning{Chi Him Ng}
\titlerunning{Enhancing Medical Learning and Reasoning Systems: A Boxology-Based Compartive Analysis of Design Patterns}

\institute{Vrije Universiteit, De Boelelaan 1105, 1081 HV Amsterdam, The Netherlands}
\maketitle              

\begin{abstract}
This study analyzes hybrid AI systems' design patterns and their effectiveness in clinical decision-making using the Boxology framework. It categorizes and compares various architectures combining machine learning and rule-based reasoning to provide insights into their structural foundations and healthcare applications. Addressing two main questions, how to categorize these systems against established design patterns and how to extract insights through comparative analysis, the study uses design patterns from software engineering to understand and optimize healthcare AI systems. Boxology helps identify commonalities and create reusable solutions, enhancing these systems' scalability, reliability, and performance. Five primary architectures are examined: REML, MLRB, RBML, RMLT, and PERML. Each has unique strengths and weaknesses, highlighting the need for tailored approaches in clinical tasks. REML excels in high-accuracy prediction for datasets with limited data; MLRB in handling large datasets and complex data integration; RBML in explainability and trustworthiness; RMLT in managing high-dimensional data; and PERML, though limited in analysis, shows promise in urgent care scenarios. The study introduces four new patterns, creates five abstract categorization patterns, and refines those five further to specific systems. These contributions enhance Boxology's taxonomical organization and offer novel approaches to integrating expert knowledge with machine learning. Boxology’s structured, modular approach offers significant advantages in developing and analyzing hybrid AI systems, revealing commonalities, and promoting reusable solutions. In conclusion, this study underscores hybrid AI systems' crucial role in advancing healthcare and Boxology's potential to drive further innovation in AI integration, ultimately improving clinical decision support and patient outcomes.


\keywords{Hybrid AI systems, Clinical decision-making, Boxology, Machine learning, Rule-based reasoning, Healthcare AI, AI architectures, Medical AI integration}
\end{abstract}

\section{Introduction}
In recent years, integrating artificial intelligence (AI) into healthcare has surged, offering opportunities for improved delivery, diagnostic accuracy, treatment planning, and patient outcomes \cite{kasula2021ai}. Systems combining learning and reasoning play a pivotal role, leveraging machine learning algorithms and reasoning mechanisms to analyze complex medical data, extract insights, and assist healthcare professionals \cite{guerrero2022decision}.

Despite the proliferation of AI-enabled medical applications, understanding and evaluating the diverse array of learning and reasoning systems remains crucial. This paper addresses two pivotal research questions to elucidate AI's role in healthcare. The first question categorizes and maps heterogeneous systems against established design patterns. Drawing from software engineering, these patterns offer a framework for analyzing healthcare AI systems' structural foundations \cite{van2021modular}. This approach reveals reusable solutions and best practices for successful deployments.

Using the Boxology framework, the study categorizes and maps hybrid AI systems in clinical decision-making, uncovering hidden commonalities and facilitating the creation of modular, reusable solutions, enhancing scalability, reliability, and performance.

The second question extracts insights through a comparative analysis of these systems within design patterns. By comparing and contrasting these systems, the study highlights the strengths and weaknesses of each design pattern. This analysis leads to specific recommendations for their application in various clinical tasks, informing the development and refinement of hybrid solutions by identifying best practices and design strategies that optimize integration with clinical knowledge. The analysis underscores the importance of tailored approaches for different medical conditions and decision-making contexts, ultimately improving the efficacy and trustworthiness of hybrid healthcare solutions.

This study builds on the systematic categorization provided in "Taxonomy of hybrid architectures involving rule-based reasoning and machine learning in clinical decision systems: A scoping review" by Kierner et al. \cite{kierner2023taxonomy}, published in the Journal of Biomedical Informatics in 2023. This paper serves as a comprehensive source of existing systems merging learning and reasoning in healthcare AI. Leveraging this examination, our methodology deepens the understanding of integrating hybrid learning and reasoning with rule-based reasoning in clinical decision systems. As AI technologies proliferate in healthcare, blending medical knowledge, often represented as clinical rules, with machine learning's predictive power is essential for creating accurate and transparent decision-making systems.

\section{Related Work}
This section provides a critical overview of the current literature and methodologies that inform the foundation of the study, focusing on advanced hybrid systems in artificial intelligence and healthcare technologies. The following two main subsections will be explored: "Boxology" for a structured approach to hybrid AI system design, and "Learning and Reasoning Systems" to highlight practical applications in healthcare. 

\subsection{Boxology}
Boxology, as presented by Van Bekkum et al., 2021, offers a systematic approach to designing complex AI systems, integrating statistical (data-driven) and symbolic (knowledge-driven) methods. The framework emphasizes a unified view of hybrid neuro-symbolic AI systems, proposing modular design patterns that describe numerous architectures using elementary building blocks. Key contributions include a taxonomically organized vocabulary, over 15 design patterns, and their application in realistic use-cases.

The framework's strength lies in revealing similarities between previously unrecognized systems through high-level architectural descriptions. It categorizes components and operations within hybrid AI systems, distinguishing between data, symbols, models, and transformations. This modular approach aligns with software engineering principles.

Boxology is preferred over other frameworks, such as those by Kautz and De Raedt et al., for its detailed taxonomy and practical design patterns. While Kautz's taxonomy lacks detailed architectural descriptions, and De Raedt et al.'s work does not address system configuration, boxology provides a more granular and practical approach. It effectively integrates statistical and symbolic methods, offering a balanced approach that leverages both paradigms' strengths.

In summary, boxology is a comprehensive framework for designing, categorizing, and analyzing hybrid AI systems. Its modular design patterns and taxonomical vocabulary offer valuable insights and tools for researchers and practitioners, enhancing the understanding and application of hybrid learning and reasoning systems.

\subsection{Learning and reasoning systems}
Numerous studies in the scientific literature have explored the utilization of systems that combine learning and reasoning in the context of healthcare, providing valuable insights into the design, implementation, and effectiveness of such systems. The following is a broad but concise overview of some approaches adopted in existing research and how they relate to and inform the present study.

One of the approaches in this domain involves the use of hybrid systems that incorporate deep learning techniques for data analysis with expert systems for decision support. Such systems have demonstrated proficiency in diagnosing diseases with high accuracy by learning from medical records and imaging data, and then applying clinical guidelines to recommend treatment plans \cite{kumar2020data}. For instance, a notable study employed a deep neural network to analyze radiographs, which was then integrated with a reasoning module that utilized medical knowledge to interpret the findings in the context of patient histories, thereby enhancing diagnostic precision and reducing false positives \cite{han2021unifying}.

Another significant area of research focuses on the use of reinforcement learning combined with rule-based reasoning for personalized medicine. In these systems, reinforcement learning algorithms optimize treatment strategies based on patient response, while reasoning mechanisms ensure adherence to clinical guidelines and patient safety protocols \cite{yu2021reinforcement}. This dual approach not only tailors treatments to individual patient needs but also incorporates the critical checks and balances necessary for safe care delivery.


The aforementioned examples highlight the diverse ways in which the combination of learning and reasoning can enhance healthcare systems. These hybrid approaches not only have the potential to improve the accuracy and efficiency of healthcare delivery but also make more room for personalized and preventive medicine. As such, they offer valuable insights into the design and implementation of future healthcare technologies.

\subsection{Framework}
This study builds on the research from Kierner et al.'s 2023 paper, "Taxonomy of hybrid architectures involving rule-based reasoning and machine learning in clinical decision systems", published in the Journal of Biomedical Informatics. Instead of a new literature review, it leverages Kierner et al.'s detailed examination and categorization of existing systems, offering insights into the integration of learning and reasoning in healthcare AI.

The study explores the integration of AI with rule-based reasoning in clinical decision systems. With AI's growing role in healthcare, blending clinical rules with ML's predictive power is essential for creating accurate and transparent decision-making systems.

Kierner et al. conducted a comprehensive literature review from 1992 to 2022, identifying and categorizing hybrid architectural patterns combining rule-based reasoning and ML. They examined articles from PubMed, IEEE Explore, and Google Scholar using keywords related to clinical decision systems and hybrid architectures, focusing on articles that utilized both approaches and detailed the architectures employed.

Through their review, the authors identified five distinct architecture archetypes used in the creation of clinical decision systems. These archetypes are as follows:

\begin{enumerate}
    \item \textbf{Rules Embedded in ML Architecture (REML):} This architecture embeds clinical rules directly into the structure of a machine learning model, such as a neural network. The rules are incorporated as part of the model's logic, enabling the system to leverage both the precision of machine learning predictions and the clarity of rule-based reasoning. \cite{kierner2023taxonomy}
    
    \item \textbf{ML Pre-processes Input Data for Rule-Based Inference (MLRB):} In this approach, a machine learning model preprocesses the input data, extracting features or making preliminary assessments, which are then fed into a rule-based system for final decision-making. This sequence allows the rule-based system to operate on more refined data, potentially improving the accuracy and relevance of its conclusions. \cite{kierner2023taxonomy}
    
    \item \textbf{Rule-Based Method Pre-processes Input Data for ML Prediction (RBML):} Contrary to MLRB, RBML starts with rule-based preprocessing of data. Here, clinical rules are applied first to organize, filter, or enhance the data before it is input into a machine learning model for the final prediction. This setup can help in reducing the complexity of the data and focusing the ML model on the most relevant features. \cite{kierner2023taxonomy}
    
    \item \textbf{Rules Influence ML Training (RMLT):} In this configuration, clinical rules are used to guide the training of the machine learning model. This might involve using rules to select or weight training data, or to adjust the model's parameters. The aim is to ensure that the machine learning model aligns more closely with established clinical knowledge and practices. \cite{kierner2023taxonomy}
    
    \item \textbf{Parallel Ensemble of Rules and ML (PERML):} This architecture runs a rule-based system and a machine learning model in parallel, with their outputs combined or adjudicated by an aggregator. This approach leverages the strengths of both systems simultaneously, offering a balance between the transparency and straightforward logic of rule-based reasoning and the analytical power of machine learning. \cite{kierner2023taxonomy}
\end{enumerate}

The study notes a potential shift towards parallel architectures (PERML), suggesting they could offer greater transparency and easier explanation, which is crucial for patient and clinician trust. Futhermore, the authors highlight a gap in the literature concerning parallel architectures in clinical settings, suggesting future research directions that could enhance the development of AI-driven solutions in healthcare.

\section{Methodology}

\subsection{Research}
As mentioned, the foundational document's taxonomy provides a framework for this study of hybrid systems combining rule-based reasoning with machine learning in healthcare AI. This exploration assesses how these systems align with established AI development patterns, aiming to understand the relationship between different learning and reasoning approaches.

In this research boxology diagrams were created for each system to visually map their structures and processes. These diagrams will show the flow of data, detailing how inputs are processed and transformed into outputs, offering a clear picture of each system's design and functionality.

Using these visualizations and extracted information from the papers, the systems were compared, analyzing their strengths, weaknesses, and unique features within healthcare applications. This comparison will help identify patterns and trends in how hybrid architectures enhance clinical decision support, showcasing their potential to improve AI-driven solutions in healthcare.

Finally, by synthesizing insights from the boxology diagrams and our comparative analysis, the research will show the strengths and weaknesses of each system and provide recommendations on which categories are best suited for specific healthcare applications. This process will deepen our understanding of how rule-based reasoning and machine learning can be effectively combined to enhance clinical decision support.

\section{Results}
This section will discuss the results across the five archetypes, REML, MLRB, RBML, RMLT, and PERML. Firstly, an overview will be provided for the clinical applications of the systems per archetype. Clinical applications are divided into four major categories: Clinical decision support, segmentation, diagnostic support, and predictive support. Clinical decision support aids healthcare providers in making clinical decisions. Segmentation involves partitioning medical images into segments for easier analysis. Diagnostic support helps in identifying diseases or conditions. Predictive support provides tools for forecasting health outcomes. 

The strengths and weaknesses of each archetype, derived from boxology diagrams and information gathered from the papers during this research, will be discussed. Furthermore, new patterns frequently appearing, not mentioned in Bekkum et al., 2021, will be discussed. \cite{van2021modular} The complete set of refined patterns is available in the appendix. In these patterns, four different steps are distinguished. Each different step is color-coded for clarity. The preprocessing steps, marked in brown, involve the preparation and cleaning of data. The category-specific steps, highlighted in pink, these include steps such as rule embedding for REML. The model creation stage, shown in blue, involves building and training the predictive models based on the preprocessed data. Finally, the output steps, indicated in green, encompass prediction, deduction, and any necessary post-processing to generate the end results. Lastly, the frequency of elementary patterns is counted per archetype. The full meaning of these patterns can be found in the paper of Bekkum et al., 2021. \cite{van2021modular}

\subsection{REML}
\begin{figure}[htbp]
    \centering
    \includegraphics[width=0.9\textwidth]{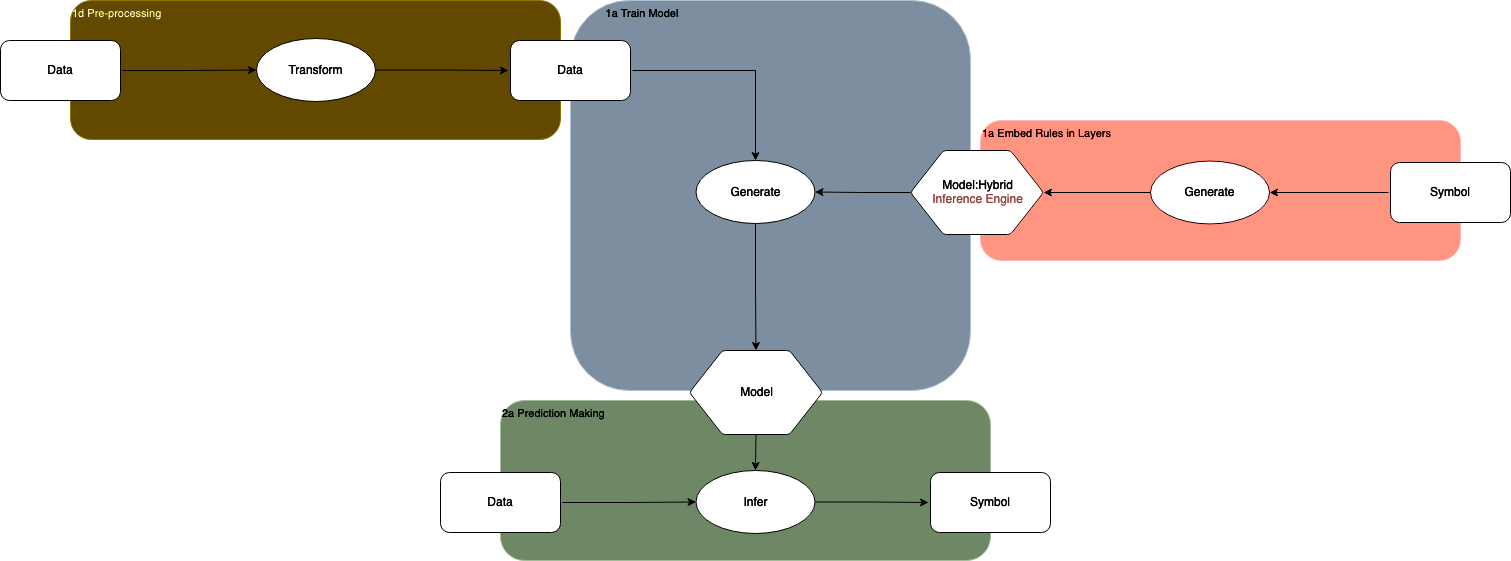}
    \caption{REML}
    \label{fig:reml}
\end{figure}

\begin{figure}[htbp]
    \centering
    \includegraphics[width=0.9\textwidth]{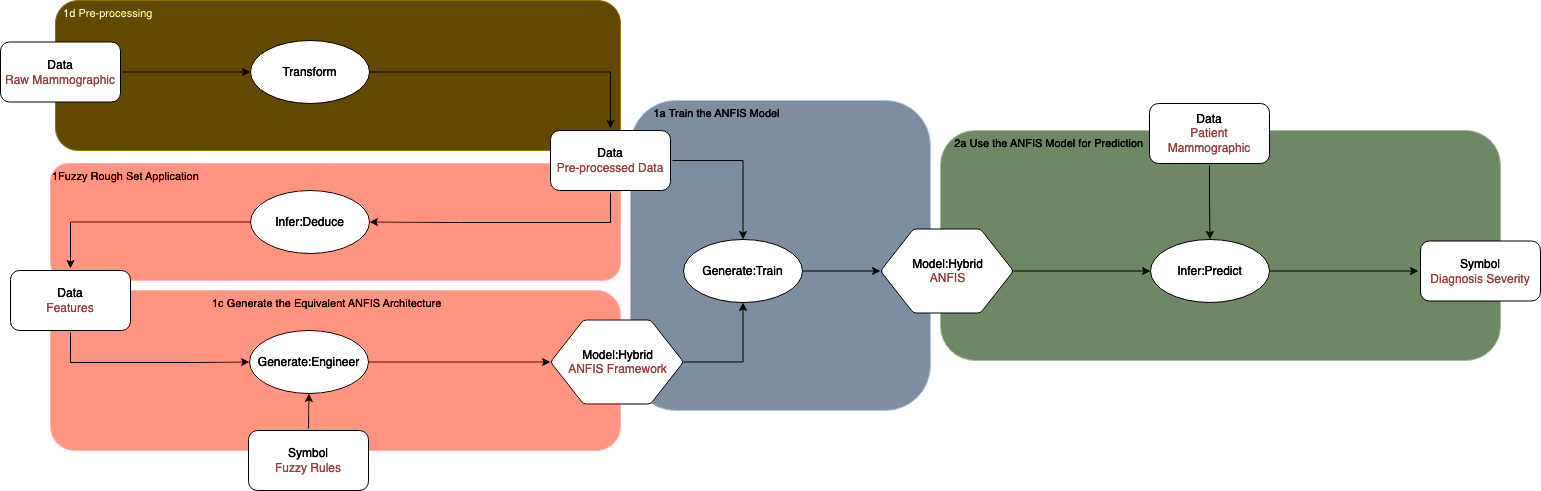}
    \caption{Usage of Case-Based Reasoning, Neural Network and Adaptive Neuro-Fuzzy Inference System Classification Techniques in Breast Cancer Dataset Classification Diagnosis}
    \label{fig:ex1_reml}
\end{figure}

\begin{figure}[htbp]
    \centering
    \includegraphics[width=0.8\textwidth]{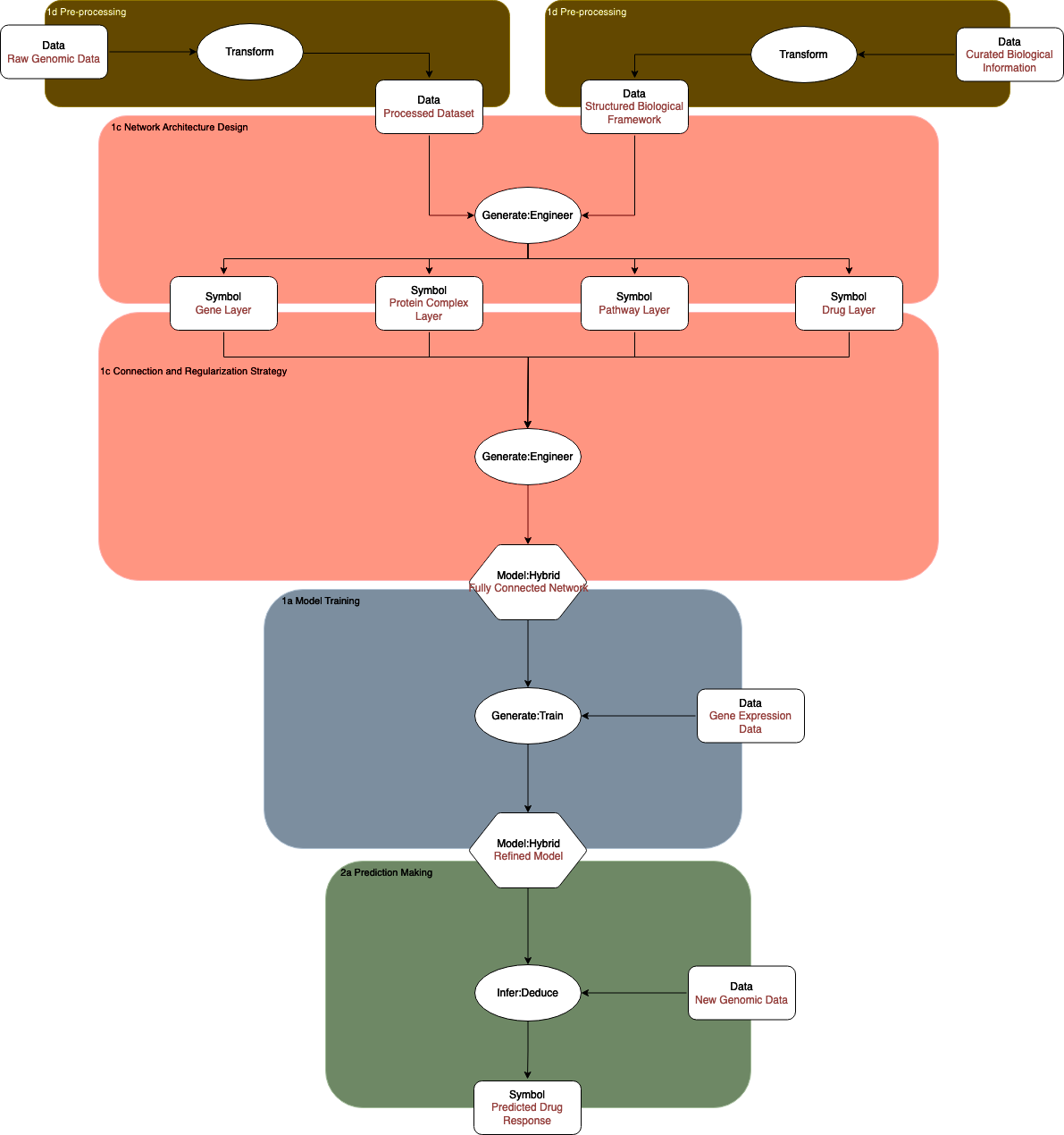}
    \caption{Biological Domain Knowledge-based Artificial Neural Network for Drug Response Prediction (BDKANN)}
    \label{fig:ex2_reml}
\end{figure}

REML systems prioritize Prediction Accuracy as their primary goal, enhancing the reliability of medical diagnoses and treatment plans by integrating sophisticated techniques like PCA and ANFIS. The secondary motivation is Data Efficiency, which focuses on reducing computational overhead by minimizing the feature set without sacrificing information quality, thus enabling quicker and more efficient predictions. The generalized design pattern of REML can be found in Figure \ref{fig:reml}. While this pattern is generalized, the embedding of rules in the model varies significantly. Two examples of different systems are shown in figures \ref{fig:ex1_reml} and \ref{fig:ex2_reml}. The Neural Network with Biological Knowledge embeds rules implicitly through a biologically informed architecture and connections, using regularization to emphasize biologically relevant pathways. In contrast, ANFIS embeds rules explicitly via predefined fuzzy logic rules and membership functions, refining them during training to combine interpretability with neural network learning. This notable difference is shown in the pink boxes. 

\subsubsection{Clinical Emphasis and Health Conditions}
In this category, 34 systems were found. Among them, five are specifically diagnostic systems. For instance, the ANFIS-Net system effectively detects COVID-19 from chest X-ray images, offering quick and precise diagnoses crucial for managing the virus \cite{al2021anfis}. For ischemic stroke patients, combining PCA and ANFIS predicts infarction volume growth rate, aiding treatment decisions \cite{ali2018use}. In lung cancer diagnosis, one system uses a neuro-fuzzy technique for detecting lung nodules in CT images, while another uses neural networks for classification, with human review ensuring reliability \cite{joshya2021automated, kuruvilla2014lung}. A knowledge-based artificial neural network enhances pulmonary embolism diagnosis by integrating rule-based reasoning with neural networks using PIOPED criteria \cite{serpen2008knowledge}.

The Adaptive Neuro-Fuzzy Inference System predicts surgery time for ischemic stroke patients by integrating multiple machine learning techniques \cite{ali2019adaptive}. The FNDSB system offers high explainability for diagnosing back pain \cite{kadhim2018fndsb}. The Causal Rule Ensemble finds heterogeneous treatment effects in environmental epidemiology, emphasizing robustness and interpretability \cite{bargagli2020causal}. Distributed Probabilistic Fuzzy Rule Mining handles large, inconsistent datasets, providing a scalable model for various clinical scenarios \cite{sharif2021distributed}.

In predictive systems, the Forced Oscillation Technique combined with machine learning and neuro-fuzzy classifiers improves differential diagnosis of asthma and restrictive respiratory diseases \cite{amaral2020differential}. A clinical decision support system uses weighted fuzzy rules to predict heart disease risk levels \cite{anooj2012clinical}. An adaptive network predicts post-dialysis urea concentrations without blood sampling, enhancing monitoring accuracy \cite{azar2013adaptive}. Cardiovascular analysis integrates autoregression and neuro-fuzzy inference systems \cite{liu2004human}. Modified AHP and Type-2 fuzzy logic enhance cancer classification accuracy \cite{nguyen2015modified}. A heart disease prediction system combines hybrid techniques with fuzzy logic to improve accuracy \cite{roy2016brain}. The BDKANN system enhances drug response predictions in cancer treatment by integrating domain knowledge \cite{snow2019bdkann}.

\begin{figure}[htbp]
    \centering
    \begin{subfigure}[b]{0.45\textwidth}
        \centering
        \includegraphics[width=\textwidth]{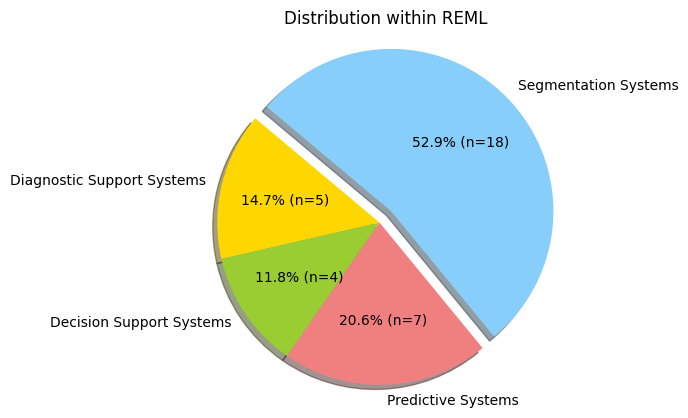}
        \caption{REML}
    \end{subfigure}
    \hfill
    \begin{subfigure}[b]{0.45\textwidth}
        \centering
        \includegraphics[width=\textwidth]{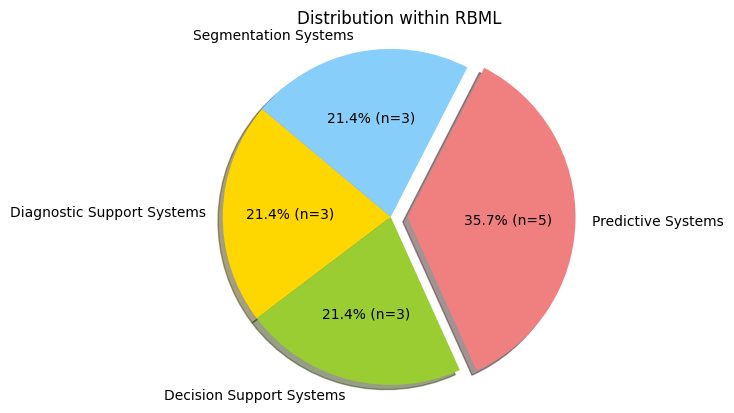}
        \caption{RBML}
    \end{subfigure}

    \vspace{0.5cm} 
    \begin{subfigure}[b]{0.45\textwidth}
        \centering
        \includegraphics[width=\textwidth]{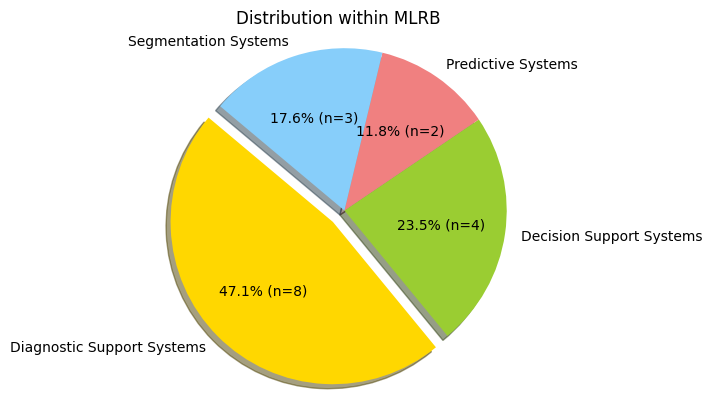}
        \caption{MLRB}
    \end{subfigure}
    \hfill
    \begin{subfigure}[b]{0.45\textwidth}
        \centering
        \includegraphics[width=\textwidth]{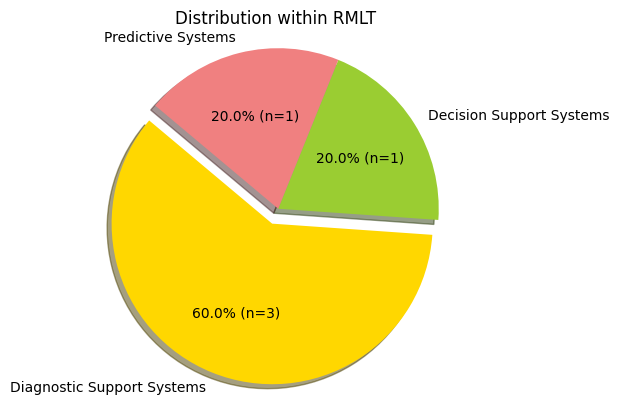}
        \caption{RMLT}
    \end{subfigure}

    \vspace{0.5cm} 

    \caption{Clinical Tasks}
    \label{fig:hybrid_architectures}
\end{figure}

\begin{figure}[htbp]
    \centering
    \includegraphics[width=0.65\textwidth]{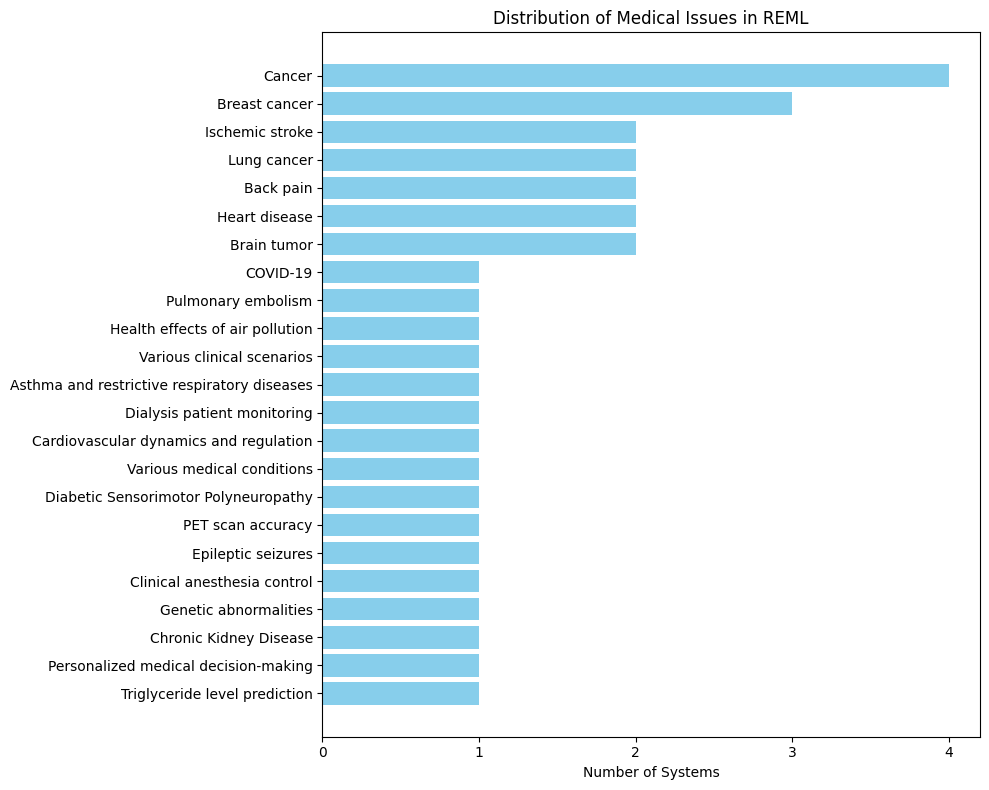}
    \caption{REML Health Conditions}
    \label{fig:reml_health}
\end{figure}

Segmentation systems include rough set theory for data analysis in various medical contexts \cite{azar2015inductive}, ANFIS for diagnosing mechanical low back pain \cite{fakharian2021diagnosis}, and rough fuzzy classifiers for accurate cancer prediction \cite{halder2019active}. ANFIS is also used in diabetic sensorimotor polyneuropathy severity classification and breast cancer dataset classification \cite{haque2021diabetic, huang2012usage}. Neural network and decision tree models predict breast cancer relapse, while fuzzy deep learning predicts lung tumor movements during radiotherapy \cite{jerez2003combined, park2016intra}. Synthetic CT generation for PET scans improves accuracy with fuzzy logic \cite{qian2019mdixon}. Epileptic seizure detection combines wavelet feature extraction with ANFIS \cite{subasi2007application}, and brain tumor detection from MRI uses ANFIS for classification \cite{thirumurugan2016brain}. Fuzzy logic is used to automate clinical anesthesia control, enhancing safety \cite{tian2022fuzzy}. Fuzzy ARTMAP efficiently classifies genetic abnormalities \cite{vigdor2006accurate}, while ANFIS detects microcalcifications in mammograms \cite{xu2007detection}. Predicting chronic kidney disease progression uses ANFIS and clinical data for accuracy \cite{yadollahpour2018designing}. ECG knowledge discovery integrates rule-based reasoning and learning classifiers for personalized decision-making \cite{zouri2020ecg}. MLP and fuzzy regression predict triglyceride levels accurately \cite{ahmad2022prediction}. Figure \ref{fig:hybrid_architectures}  presents a detailed chart categorizing various systems, with a notable emphasis on segmentation tasks. This highlights a significant trend within the domain towards segmentation-focused solutions. To further enhance understanding, Figure \ref{fig:reml_health} illustrates the frequency of health conditions addressed by these systems. It is apparent that there is a predominant focus on cancer-related applications. Together, these figures provide a comprehensive overview of the current landscape, underscoring the primary areas of interest and application within the field.

\subsubsection{Strengths of REML Systems}
The most important strengths of REML systems are their high diagnostic and predictive accuracy, efficiency with small datasets, and ability to handle uncertainty. Systems like ANFIS-Net and those designed for stroke, asthma, and heart disease achieve exceptional accuracy rates, significantly enhancing their reliability and utility in clinical settings \cite{al2021anfis, ali2019adaptive, amaral2020differential}. These systems, such as those for epileptic seizure detection and mammogram analysis, maintain high performance even with limited data, which makes them highly practical for various medical applications where data might be scarce or difficult to obtain \cite{subasi2007application, vigdor2006accurate}. The integration of fuzzy logic enables these systems to manage imprecision and nonlinear relationships effectively, thereby improving their decision-making capabilities and handling the inherent uncertainty in medical data \cite{azar2013adaptive, halder2019active, haque2021diabetic}.

Additionally, REML systems exhibit a high degree of adaptability, efficiently learning from new data and adjusting to evolving information, which is crucial for maintaining their relevance and accuracy over time \cite{al2021anfis, halder2019active}. Their transparency through rule-based components provides clear decision-making processes, which is vital for gaining trust and acceptance among clinicians. This transparency is exemplified in systems like the neuro-fuzzy classifier for asthma diagnosis and heart disease risk prediction, where understandable decision-making processes enhance clinical trust and usability \cite{amaral2020differential, anooj2012clinical}. Furthermore, these systems have demonstrated their adaptability by effectively learning from new information and evolving data environments, as seen in the DSPN severity classification and breast cancer diagnosis systems \cite{halder2019active, haque2021diabetic}. This adaptability, combined with their high diagnostic accuracy, efficiency with small datasets, underscores the overall effectiveness and clinical acceptance of REML systems in various medical applications.

\subsubsection{Weaknesses of REML Systems}
\begin{figure}[htbp]
    \centering
    \includegraphics[width=0.9\textwidth]{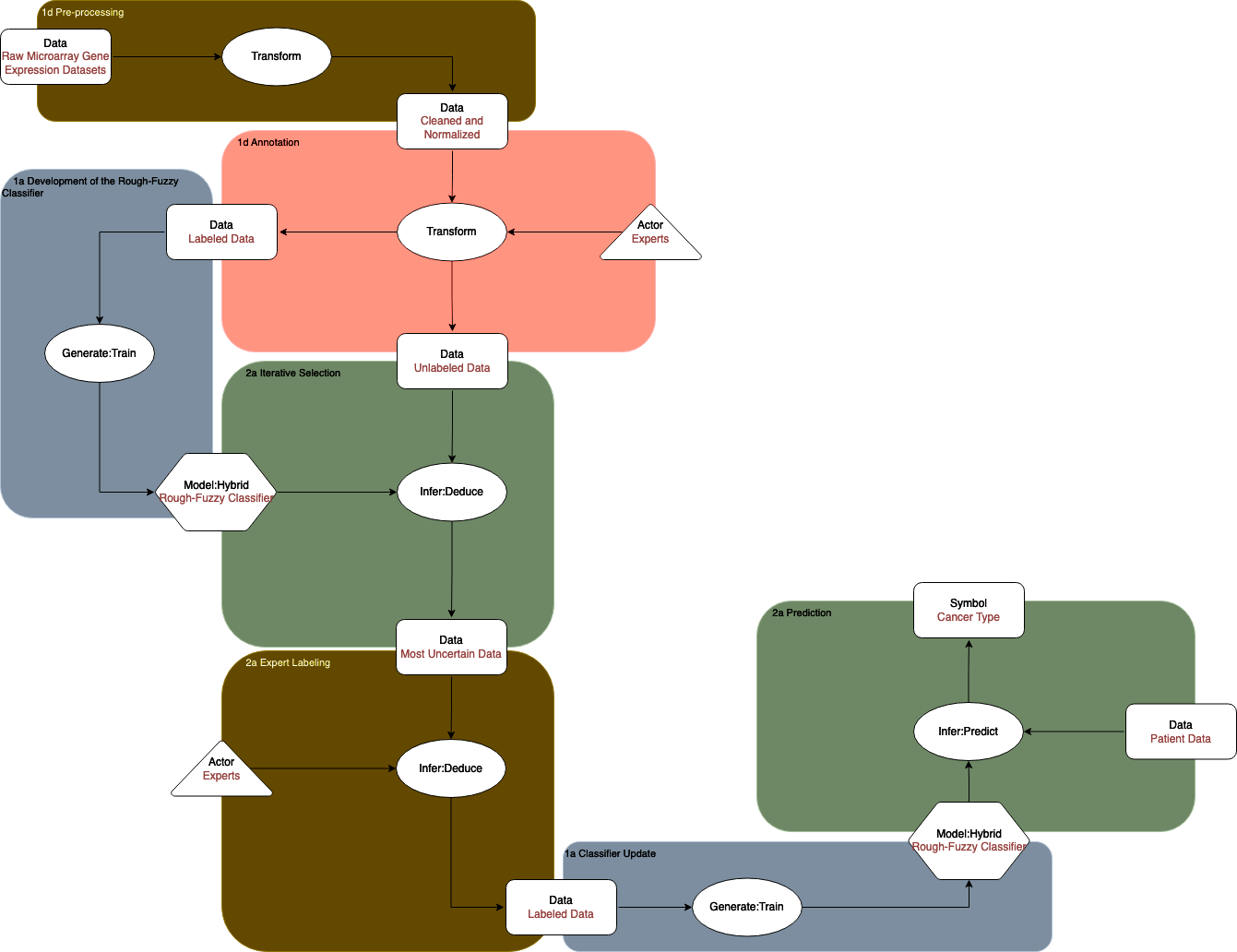}
    \caption{Active learning using rough fuzzy classifier for cancer prediction from microarray gene expression data}
    \label{fig:ex3_reml}
\end{figure}

The shortcomings of REML systems present considerable obstacles to their use in a clinical setting. A significant challenge is the intricacy involved in optimization and implementation, since REML is heavily depended on embedded rules, which require the adjustment of multiple parameters and a high level of technical expertise. This complexity can slow down their uptake and effective use in healthcare \cite{al2021anfis, ali2018use, azar2015inductive}. Additionally, many systems heavily rely on cleaned input data, which influences their accuracy and performance. For instance, ANFIS-Net depends on effective chest X-ray feature extraction, and the quality of FOT measurements impacts the respiratory disease diagnosis system \cite{al2021anfis, amaral2020differential, fakharian2021diagnosis}. Such example can be seen in Figure \ref{fig:ex3_reml}. The system heavily relies on cleaned rules and data because the accuracy of fuzzy membership calculations and rough set approximations directly depends on the quality of the initial labeled dataset and the precision of the rules derived from it. Additionally, the processes of generating fuzzy parameters and labeling informative samples require human expertise. Any errors or noise in the data can propagate through the system, leading to incorrect classifications.

Another major issue is the limited validation across various populations and settings, which can cast doubt on their generalizability. Systems like the surgery time prediction for ischemic stroke patients and the respiratory disease diagnosis system have not been extensively tested outside specific populations \cite{ali2019adaptive, amaral2020differential}. The computational demands and significant resources needed also hinder the use of complex systems like neuro-fuzzy classifiers in real-time or resource-limited situations \cite{amaral2020differential, fakharian2021diagnosis, halder2019active}.

Other challenges include the need for human input and constant updates, which can be difficult to maintain. Systems predicting infarction volume growth rate require ongoing validation to keep pace with evolving medical practices \cite{ali2018use, ali2019adaptive, azar2015inductive}. Additionally, systems such as the Causal Rule Ensemble for estimating heterogeneous treatment effects and the mDixon-Based Synthetic CT Generation system are complex and resource-intensive, requiring advanced statistical techniques and significant computational time \cite{bargagli2020causal, park2016intra}.

Moreover, several systems face significant computational demands and depend on high-quality data for accurate classifications. For example, the Genetic Fuzzy Logic System for breast cancer and heart disease diagnosis, and the Brain Tumor Classification using ANFIS, require substantial computational resources and high-quality input data \cite{nguyen2015classification, roy2016brain}. Systems like the Knowledge-Based Artificial Neural Network for pulmonary embolism diagnosis and the Distributed Probabilistic Fuzzy Rule Mining system for clinical decision-making also face challenges related to tuning, optimization, and maintenance, which can complicate their implementation in clinical settings \cite{serpen2008knowledge, sharif2021distributed}.

\subsection{MLRB}
\begin{figure}[htbp]
    \centering
    \includegraphics[width=0.9\textwidth]{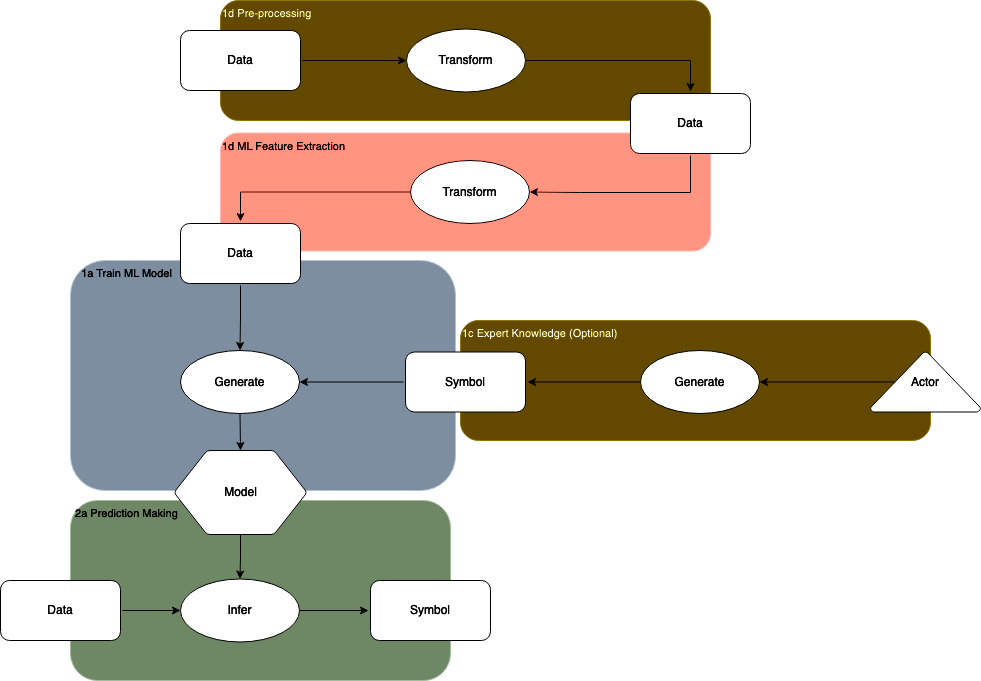}
    \caption{MLRB}
    \label{fig:mlrb}
\end{figure}

\begin{figure}[htbp]
    \centering
    \includegraphics[width=0.9\textwidth]{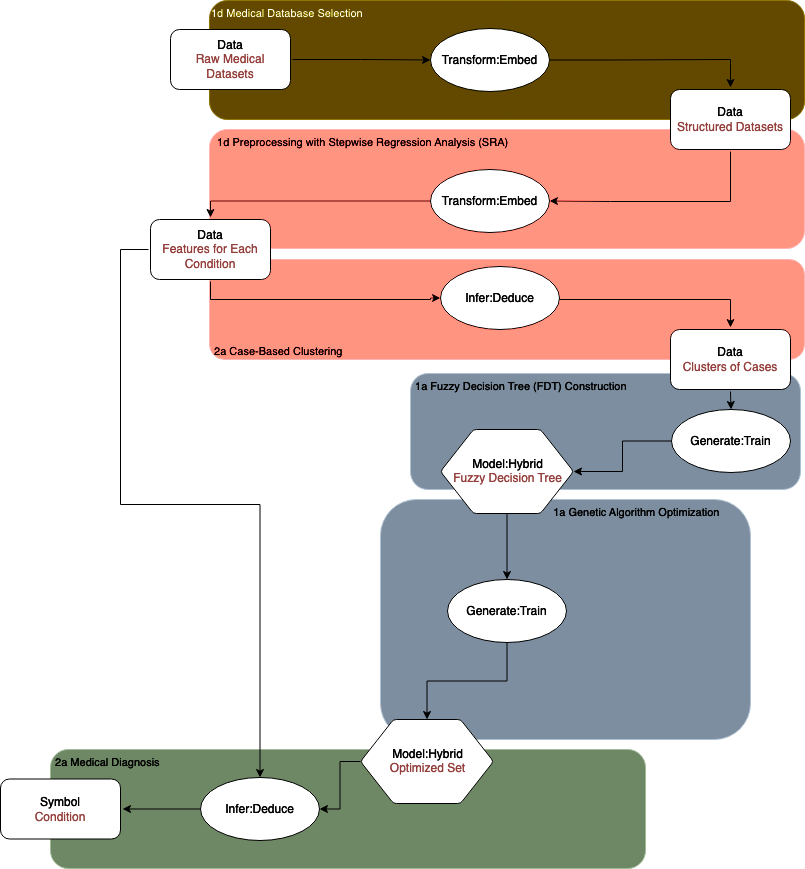}
    \caption{A hybrid model combining case-based reasoning and fuzzy decision tree for medical data classification}
    \label{fig:ex1_mlrb}
\end{figure}

\begin{figure}[htbp]
    \centering
    \includegraphics[width=0.9\textwidth]{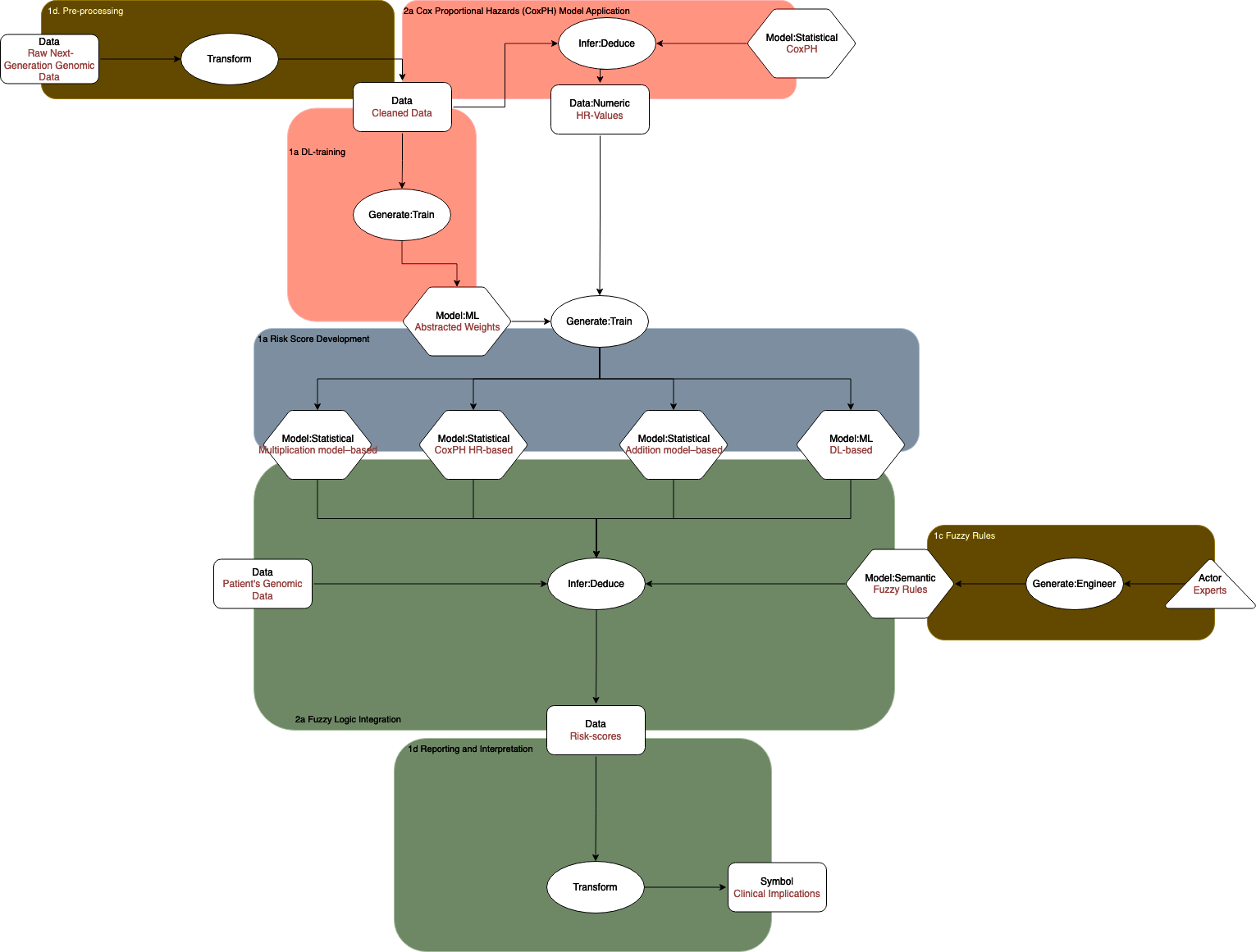}
    \caption{Applications of Deep Learning and Fuzzy Systems to Detect Cancer Mortality in Next-Generation Genomic Data}
    \label{fig:ex2_mlrb}
\end{figure}

MLRB systems prioritize Prediction Accuracy as their primary goal. The secondary motivation is Data Efficiency, these systems are designed to maximize the utility of available data. The generalized design pattern of MLRB can be found in Figure \ref{fig:mlrb}. The pre-processing methods us ML-methods varies widely. As can be seen in figures \ref{fig:ex1_mlrb} and \ref{fig:ex2_mlrb}. The two systems differ in their preprocessing stages using ML techniques. The first system focuses on preparing genomic and clinical data for dual modeling with CoxPH and deep learning. It primarily involves cleaning and preprocessing data to ensure quality, followed by using the CoxPH model to estimate hazard ratios and a deep learning model to highlight feature importance. This results in comprehensive risk scores for cancer mortality. In contrast, the second system employs more advanced techniques such as stepwise regression for feature selection and case-based clustering to simplify the dataset. After selecting significant features, the data is clustered into homogeneous groups. Fuzzy decision trees are then constructed and optimized using genetic algorithms, leading to precise and interpretable diagnostic rules for medical diagnosis.

\subsubsection{Clinical Emphasis and Health Conditions}
A total of 17 MLRB systems were identified. For diagnosis, the Ayaresa App uses SVM/NLP and fuzzy logic for general diagnosis in rural Ghana, focusing on early diagnosis, accessibility, and efficiency in underserved regions \cite{afoakwa2021android}. Similarly, SEDMAS-RBC-Adapt diagnoses acute bacterial meningitis with improved learning capabilities for precise and efficient results \cite{cabrera2010integration}. HIROFILOS-II covers prostate diseases using hybrid intelligence and fuzzy rule extraction, while a system for diagnosing gastrointestinal cancer combines case-based and rule-based reasoning, validated with real patient data \cite{koutsojannis2009using, saraiva2016early}.

In clinical decision support, systems like Clinical DSS for Nosocomial Infections integrate expert knowledge and dynamic management to improve ICU infection control. Sinedie for Gestational Diabetes automates analysis and therapy suggestions, vital for effective management \cite{caballero2017web, benomrane2014towards}. The DS Diagnostic Support system emphasizes personalized decision-making, prioritizing accuracy, interpretability, and personalization \cite{valente2021improving}. The CCSO-SOFLP System classifies pain perception from fMRI data with computational efficiency and adaptability \cite{anter2020new}. Systems for epilepsy classification from EEG signals and wearable sensors highlight advanced technologies to enhance prediction accuracy and model interpretability using expert-created fuzzy rules \cite{jiang2017seizure, kadu2022novel}.

Predictive and forecasting systems for COVID-19 trends use innovative hybrid approaches for accurate forecasts, critical for pandemic management \cite{castillo2020forecasting}. The FuzzyDeepCoxPH system identifies high-risk cancer mutations, offering vital insights for prognosis \cite{yang2020applications}.

\begin{figure}[htbp]
    \centering
    \begin{minipage}{0.48\textwidth}
        \centering
        \includegraphics[width=\textwidth]{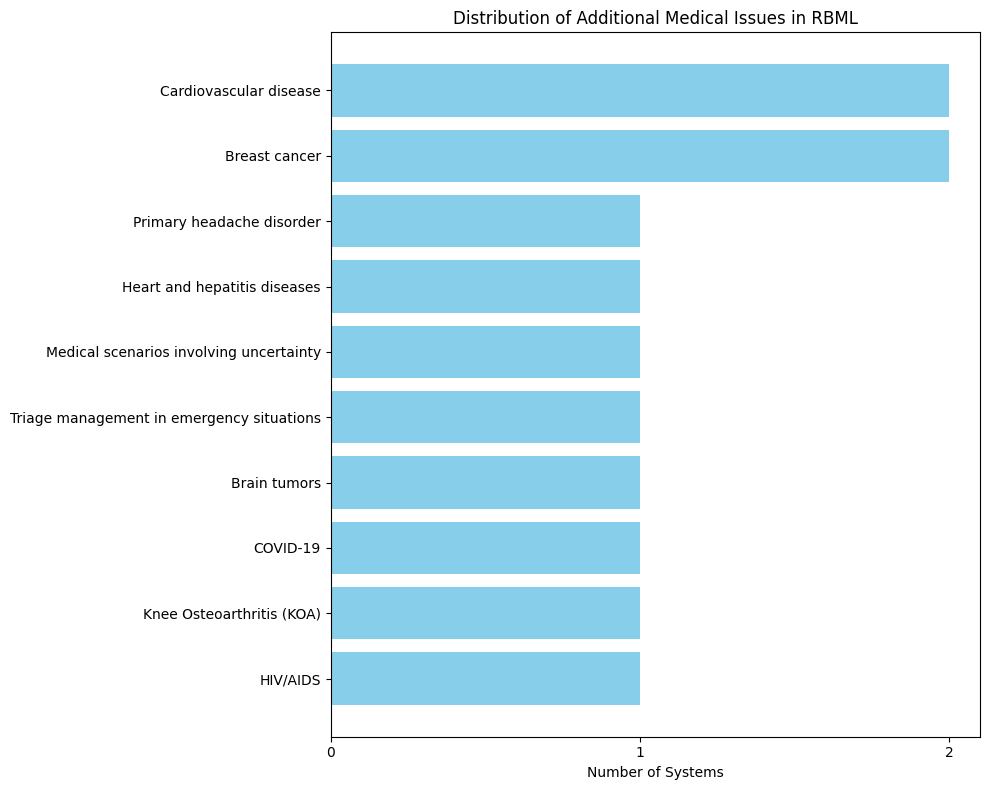}
        \caption{RBML Health Conditions}
        \label{fig:rbml_health}
    \end{minipage}\hfill
    \begin{minipage}{0.48\textwidth}
        \centering
        \includegraphics[width=\textwidth]{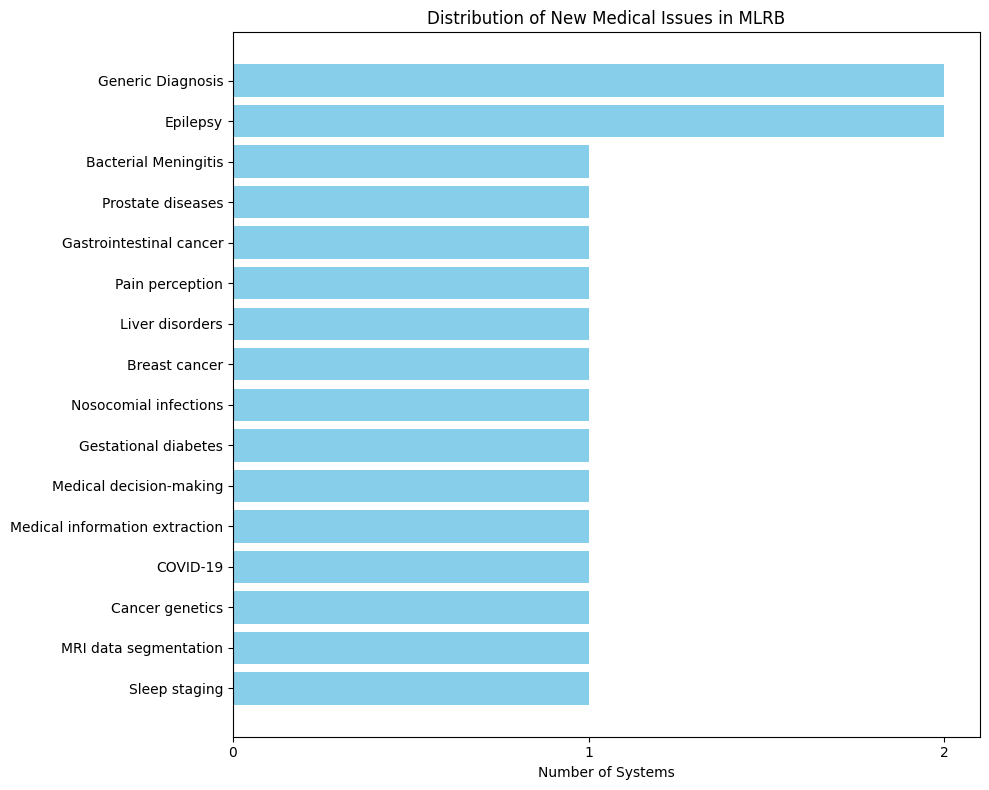}
        \caption{MLRB Health Conditions}
        \label{fig:mlrb_health}
    \end{minipage}
\end{figure}

Additionally, the Fuzzy Markov Random Field Model and Automated Personalized Sleep Staging system demonstrate sophisticated data segmentation and personalization capabilities, improving diagnostic precision and adaptability \cite{chen2018towards, daniels2017unsupervised}. To give a better overview a chart for the different categories can be seen in figure \ref{fig:hybrid_architectures}. This distribution highlights that nearly half of the systems are focused on providing diagnostic support.
Lastly, the frequency of health conditions are shown in figure \ref{fig:mlrb_health}. Generic Diagnosis and Epilepsy are the most frequently addressed issues.
 
\subsubsection{Strengths of MLRB Systems}
By enabling early diagnosis and treatment, these systems improve patient outcomes. Integrating technologies like NLP and fuzzy logic, they efficiently process conversational inputs, making it easier for users to accurately report symptoms. This enhances both user experience and diagnostic accuracy, expanding healthcare reach in both urban and rural settings. \cite{afoakwa2021android, cabrera2010integration}

In computational efficiency, systems like CCSO-SOFLP excel at managing large, complex datasets typical in areas like brain imaging. This capability is crucial for quick and accurate data processing, supporting timely medical interventions. \cite{anter2020new} This strength is often mentioned, several systems are using advanced methodologies like neural fuzzy systems and dynamic ontologies, they optimize based on data patterns, enhancing their ability to process complex datasets and extract actionable insights. \cite{baig2016machine, yang2020applications,valente2021improving}

Furthermore, other systems combine case-based reasoning and fuzzy logic. This integration boosts prediction accuracy and robustness, critical in settings where diagnostic precision is paramount. \cite{castillo2020forecasting, chen2018towards, daniels2017unsupervised}

Lastly, their adaptability and continuous learning capabilities allow MLRB systems to evolve with new medical knowledge and practices, maintaining relevance and effectiveness in a rapidly changing healthcare landscape. \cite{caballero2017web, kadu2022novel} 

\subsubsection{Weaknesses of MLRB Systems}
A notable drawback of MLRB systems is the dependence on the accuracy of user inputs. For example, if symptom reporting isn't accurate, it can lead to misdiagnoses, which in turn can lead to inappropriate treatments or unnecessary tests. This underscores the need for medical specialist involvement and highlights the limitations of these systems in managing a wide variety of diseases and symptoms. \cite{afoakwa2021android, cabrera2010integration}

Moreover, the continual evolution of medical knowledge can make it challenging to manage and update rules and methodologies. This complexity could potentially limit the systems' ability to adapt to new medical conditions or updated healthcare protocols, which could affect their effectiveness in dynamic clinical environments. \cite{anter2020new, baig2016machine}

The performance of MLRB systems also faces constraints due to the inherent complexity and incompleteness of clinical data. Despite their advanced processing capabilities, these systems rely on predefined diagnostic rules, which means their performance is heavily dependent on the completeness and accuracy of these rules. The rules may not always capture the full complexity of certain conditions. \cite{baig2016machine, valente2021improving}

Dealing with the non-linearity and high-dimensionality of data is another significant challenge. These characteristics are especially prevalent in areas like genomics, requiring specific architectural and methodological approaches. This adds to the challenge of achieving efficient data processing and accurate results. \cite{anter2020new, yang2020applications}

Lastly, another weakness mentioned multiple times are the computational demands of these systems. This makes them resource-intensive and can limit their scalability, adding to the challenges of maintaining and upgrading these systems in a clinical context. \cite{caballero2017web, castillo2020forecasting, daniels2017unsupervised}

\subsection{RBML}
\begin{figure}[htbp]
    \centering
    \includegraphics[width=0.9\textwidth]{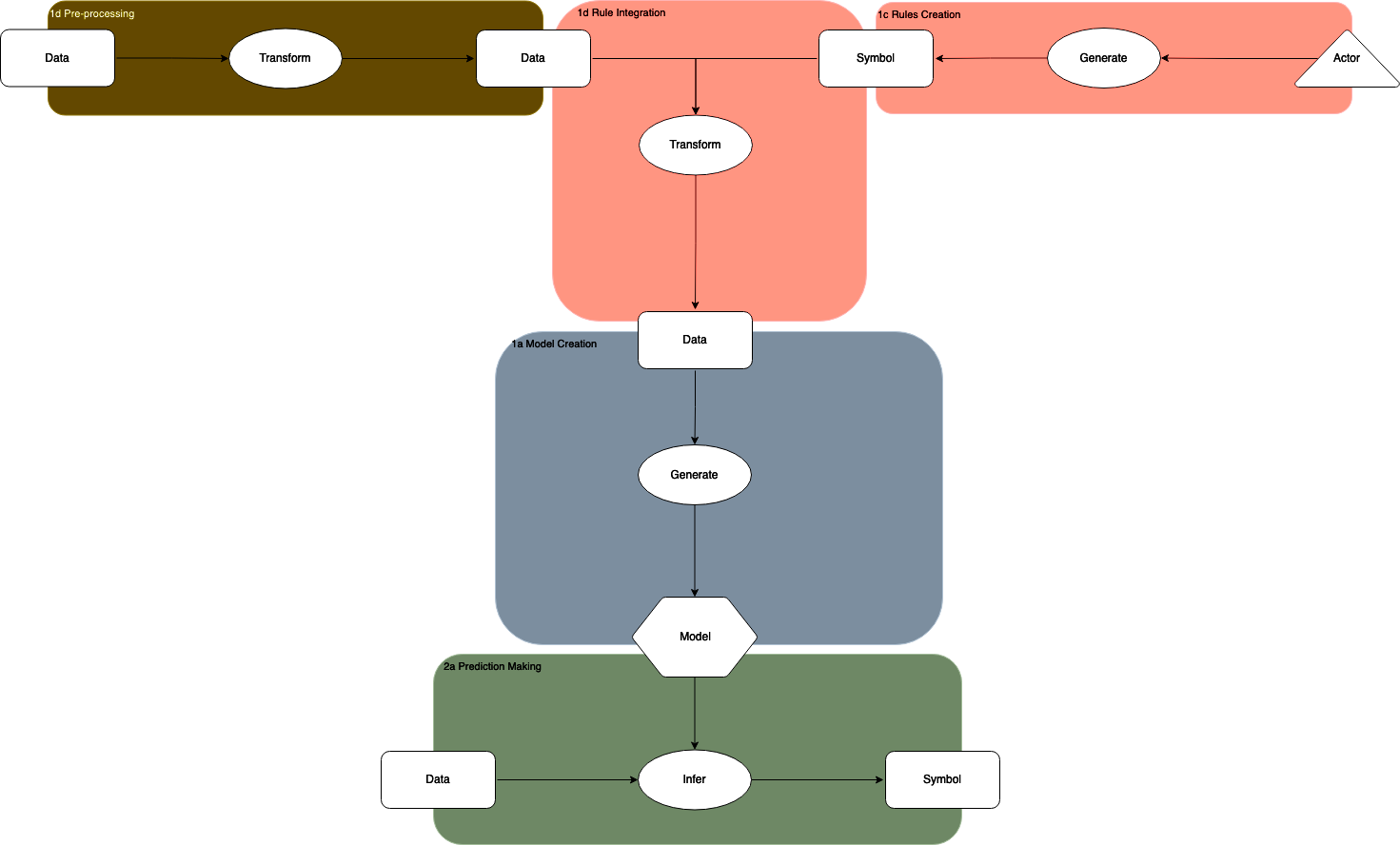}
    \caption{RBML}
    \label{fig:rbml}
\end{figure}

RBML systems prioritize Prediction Accuracy as their primary goal, ensuring reliable and accurate diagnoses by integrating rule-based methods with machine learning. The secondary motivation is Explainability and Trustworthiness, essential for providing clear reasoning behind predictions, building trust among healthcare professionals, and facilitating easier interpretation and validation of the system's recommendations. The generalized design pattern of RBML can be found in Figure \ref{fig:rbml}.

\begin{figure}[htbp]
    \centering
    \includegraphics[width=0.9\textwidth]{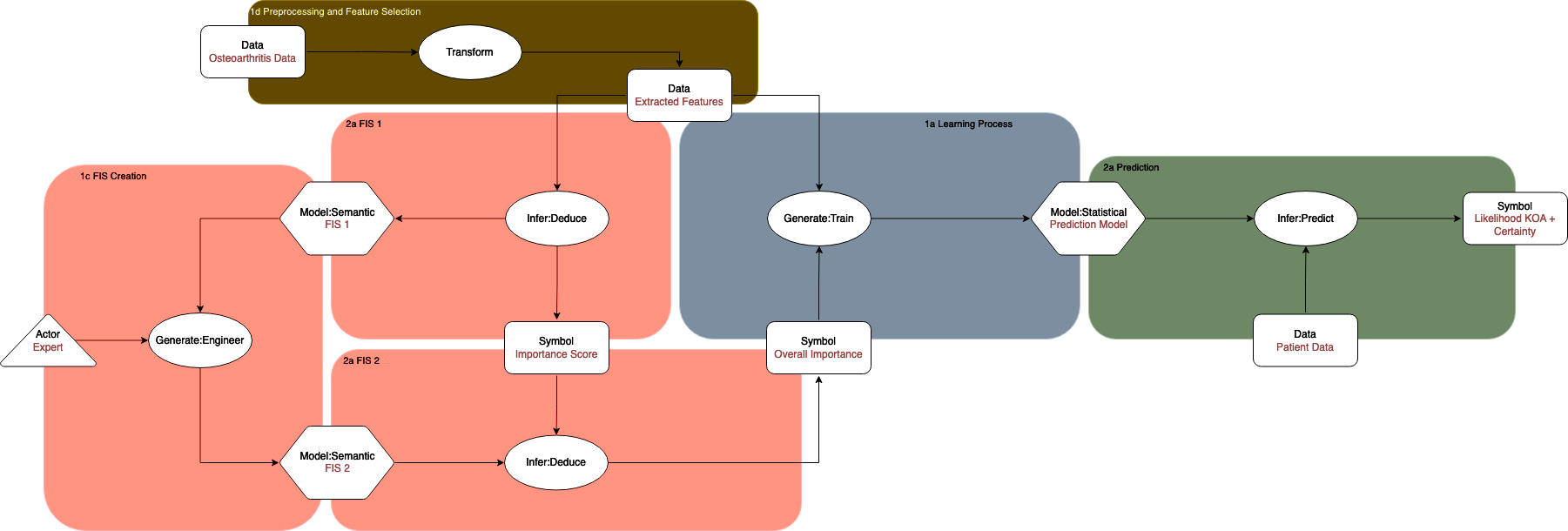}
    \caption{Explainable machine learning for knee osteoarthritis diagnosis based on a novel fuzzy feature selection methodology}
    \label{fig:ex2_rbml}
\end{figure}

\begin{figure}[htbp]
    \centering
    \includegraphics[width=0.9\textwidth]{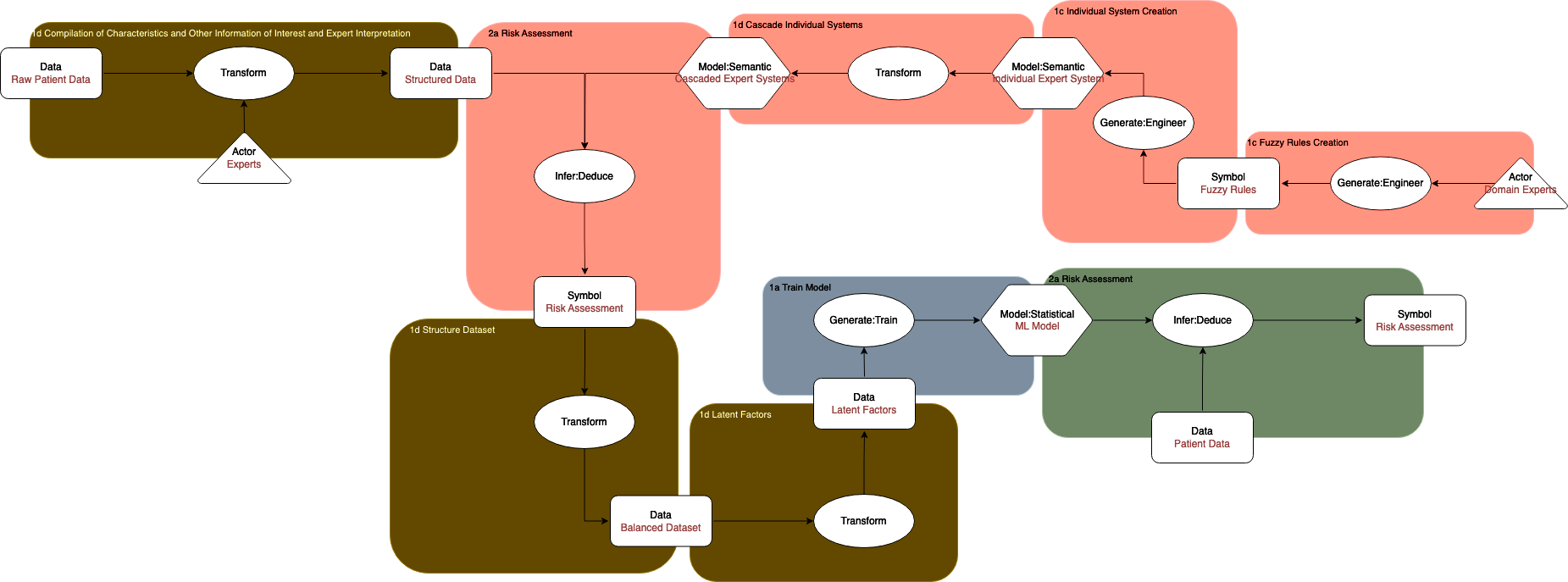}
    \caption{Design and Development of an Intelligent Clinical Decision Support System Applied to the Evaluation of Breast Cancer Risk}
    \label{fig:ex3_rbml}
\end{figure}

The method of rule-based processing can differ as can be seen in figures \ref{fig:ex2_rbml} and \ref{fig:ex3_rbml}. In the first system, rules are applied during feature selection and explainability analysis. Data is first normalized, balanced, and cleaned, then important features are selected using a method based on importance scores. After training multiple machine learning models, SHAP values are used to interpret the best model's decision-making process. In contrast, the second system uses expert systems and fuzzy logic early in the process to evaluate patient data, generating risk values for various breast cancer indicators. These values are normalized and balanced, and exploratory factorial analysis (EFA) is used to identify latent factors, reducing dimensionality. The resulting factors are used to train a machine learning model to predict breast cancer risk, leading to a risk assessment and recommendations for further medical action.

\subsubsection{Clinical Emphasis and Health Conditions}
The Hierarchical Fuzzy System for Risk Assessment of Cardiovascular Disease uses fuzzy logic for non-invasive risk assessments, enhancing preventive care with early interventions \cite{casalino2020hierarchical}. Similarly, the Hybrid Fuzzy Clustering Approach for Diagnosing Primary Headache Disorder improves diagnostic specificity for primary headaches using advanced clustering algorithms \cite{simic2021hybrid}. The Intelligent Clinical Decision Support System for Breast Cancer Risk leverages expert systems and machine learning for early detection strategies \cite{casal2022design}.

In Clinical Decision Support Systems, the Hybrid Approach to Medical Decision Support Systems combines feature selection, fuzzy weighted preprocessing, and AIRS for accurate heart and hepatitis disease diagnosis \cite{polat2007hybrid}. The Hybrid CBR and BN Architecture integrates Bayesian Networks with Case-Based Reasoning to enhance personalized medicine, such as predicting pain levels in cancer patients \cite{bruland2011hybrid}. The Decision Support System for Triage Management improves emergency patient assessments using fuzzy logic and rule-based reasoning \cite{soufi2018decision}.

Predictive and Treatment Systems include the Novel Ontology and Machine Learning Inspired Hybrid Cardiovascular Decision Support Framework, which merges ontology-driven assessments with machine learning to optimize patient referrals and preventive measures \cite{hussain2015novel}. The Efficient Approach for Brain Tumor Detection uses fuzzy logic and U-NET CNN classification for precise tumor localization from MRI scans \cite{maqsood2021efficient}. The COVID-19 Risk Prediction for Diabetic Patients system employs a fuzzy inference system with machine learning to manage risks and provide timely responses \cite{aggarwal2022covid}. Explainable Machine Learning for Knee Osteoarthritis Diagnosis applies a fuzzy feature selection methodology, enhancing model transparency and clinician trust \cite{kokkotis2022explainable}. The Self-Learning Fuzzy Discrete Event System for HIV/AIDS Treatment adapts to changing guidelines using expert feedback to ensure accurate treatment recommendations \cite{ying2007self}.

In Segmentation Systems, the Automatic Semantic Segmentation of Breast Tumors in Ultrasound Images combines fuzzy logic with deep learning to improve tumor boundary identification accuracy \cite{badawy2021automatic}. To give a better overview, the frequency of health conditions is shown in Figure \ref{fig:rbml_health}. Lastly, a chart for the different categories can be seen in figure \ref{fig:hybrid_architectures}. As shown, the RBML systems focuses on critical medical issues like cardiovascular disease and breast cancer, with a balanced distribution of system functionalities, emphasizing predictive capabilities. 

\subsubsection{Strengths of RBML Systems}
The most common strengths of RBML systems are improved prediction accuracy, enhanced explainability, and effective handling of uncertainty. Systems like the Hybrid Approach to Medical Decision Support and the Efficient Approach for Brain Tumor Detection achieve high classification accuracy, making them reliable in clinical applications. For example, the Hybrid Approach to Medical Decision Support Systems has achieved high classification accuracy for heart (92.59\%) and hepatitis diseases (81.82\%) \cite{polat2007hybrid}. Similarly, the Efficient Approach for Brain Tumor Detection using Fuzzy Logic and U-NET CNN Classification aids early diagnosis with high accuracy \cite{maqsood2021efficient}.

Enhanced explainability is another significant strength. Techniques like hierarchical fuzzy logic and SHapley Additive exPlanations (SHAP) improve model interpretability, making it easier for clinicians to trust system outputs. For instance, the Hierarchical Fuzzy System for Risk Assessment of Cardiovascular Disease reduces the number of rules, enhancing interpretability for clinicians \cite{casalino2020hierarchical}. The Explainable Machine Learning for Knee Osteoarthritis Diagnosis system uses SHAP to ensure clinician comprehension and trust \cite{kokkotis2022explainable}.

Effective handling of uncertainty is crucial for RBML systems. By integrating fuzzy logic, these systems manage imprecision in medical data, vital for accurate diagnoses. The Hybrid Fuzzy Clustering Approach for Diagnosing Primary Headache Disorder handles data variability and uncertainty, ensuring more accurate diagnostic outcomes \cite{simic2021hybrid}.

Other strengths include flexibility and integration of expert knowledge. Systems like the Self-Learning Fuzzy Discrete Event System for HIV/AIDS Treatment Regimen Selection adapt to new data and expert decisions, maintaining accuracy \cite{ying2007self}. The Hybrid CBR and BN Architecture combines Bayesian Networks and Case-Based Reasoning to enhance prediction accuracy and ground recommendations in clinical experience \cite{bruland2011hybrid}. These strengths contribute to the clinical acceptance of RBML systems, supported by advanced techniques and expert knowledge to enhance accuracy and interpretability in various clinical applications \cite{kuruvilla2014lung, nguyen2015modified, roy2016brain, serpen2008knowledge, sharif2021distributed}.

\subsubsection{Weaknesses of RBML Systems}
RBML systems have several weaknesses, including complexity in integration and maintenance, dependence on data quality, potential for overfitting, and challenges in scalability and real-world application. One significant weakness is complexity in integration and maintenance. For example, the Hierarchical Fuzzy System for Cardiovascular Disease Risk Assessment requires significant effort in rule definition and refinement, making it challenging to maintain and update, which hinders its long-term usability in clinical settings \cite{casalino2020hierarchical}. Many systems experience similar issues \cite{badawy2021automatic, bruland2011hybrid, hussain2015novel, casal2022design, ying2007self}.

Another major weakness is the heavy reliance on data quality. The Hybrid Approach to Medical Decision Support Systems performs well with high-quality data but is significantly affected by inaccuracies or missing values, limiting its effectiveness in real-world settings where data can be incomplete or noisy \cite{polat2007hybrid}.

Scalability issues also pose significant challenges. The Self-Learning Fuzzy Discrete Event System for HIV/AIDS Treatment Regimen Selection, while adaptable, may struggle to scale effectively, especially in resource-constrained environments that lack continuous expert input, limiting its applicability in diverse healthcare settings \cite{ying2007self}.

Other weaknesses include high computational demands and difficulties in model generalization. For instance, the Efficient Approach for Brain Tumor Detection using Fuzzy Logic and U-NET CNN Classification increases computational complexity and processing time, requiring substantial computing resources, limiting its real-time application \cite{maqsood2021efficient}. The Explainable Machine Learning for Knee Osteoarthritis Diagnosis faces a trade-off between interpretability and accuracy. Integrating multiple feature selection algorithms, while improving interpretability, increases computational demands, impacting scalability and making it challenging to maintain high accuracy while ensuring explainability \cite{kokkotis2022explainable}. Additionally, the Decision Support System for Triage Management and the COVID-19 Risk Prediction System for Diabetic Patients show high accuracy but may struggle to generalize rules to different clinical scenarios, necessitating significant adaptation for diverse patient populations \cite{soufi2018decision, aggarwal2022covid}.

\subsection{RMLT}
\begin{figure}[htbp]
    \centering
    \includegraphics[width=0.9\textwidth]{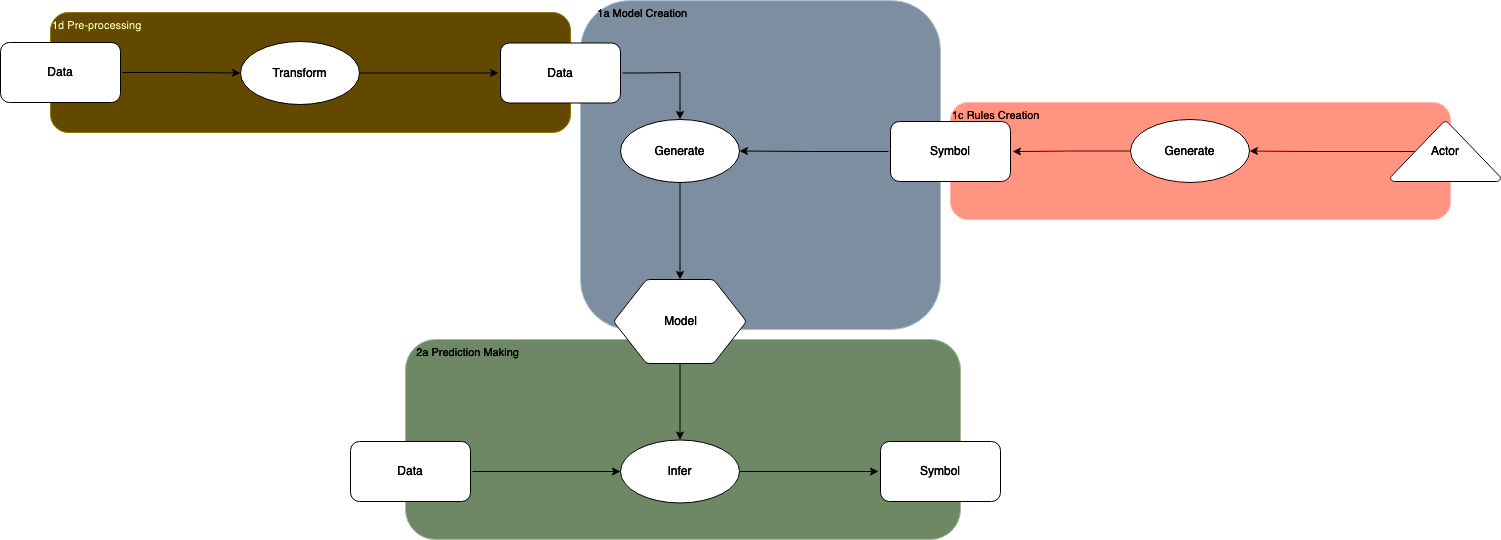}
    \caption{RMLT}
    \label{fig:rmlt}
\end{figure}

\begin{figure}[htbp]
    \centering
    \includegraphics[width=0.65\textwidth]{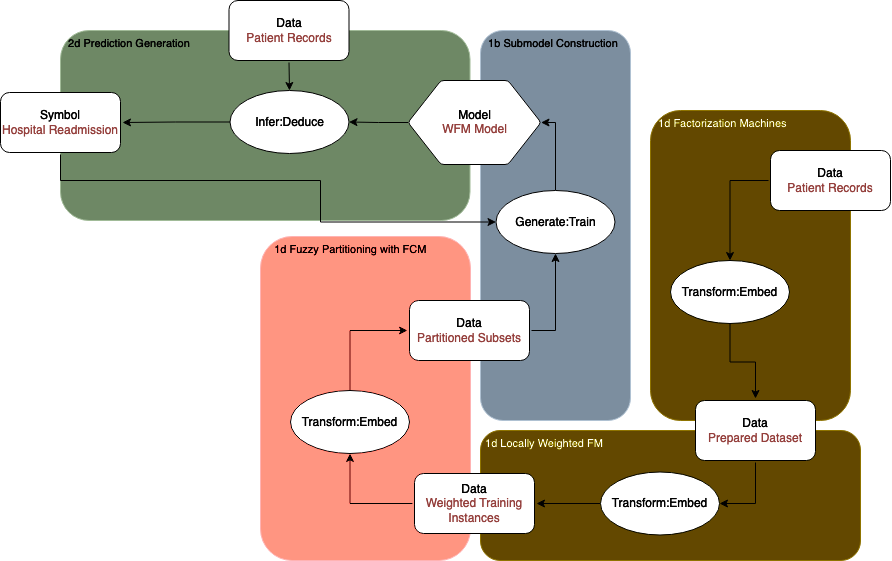}
    \caption{Locally weighted factorization machine with fuzzy partition for elderly readmission prediction}
    \label{fig:ex1_rmlt}
\end{figure}

\begin{figure}[htbp]
    \centering
    \includegraphics[width=0.75\textwidth]{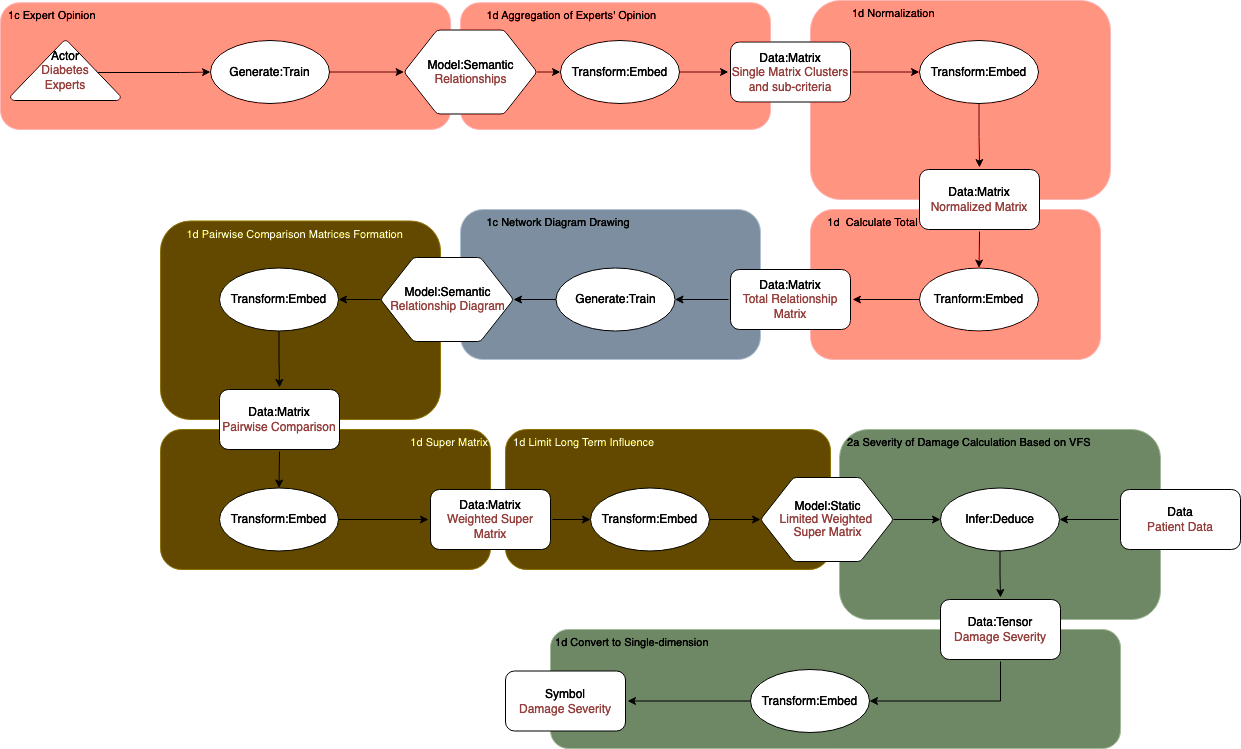}
    \caption{A novel algorithm based on information diffusion and fuzzy MADM methods for analysis of damages caused by diabetes crisis}
    \label{fig:ex2_rmlt}
\end{figure}

RMLT systems prioritize Prediction Accuracy as their primary goal. The secondary motivation revolves around Data Efficiency. The generalized design pattern of RMLT can be found in Figure \ref{fig:rmlt}. Figures \ref{fig:ex1_rmlt} and \ref{fig:ex2_rmlt} show two different systems. The key difference is that the first system uses a data-driven approach where clustering and local weighting guide the training and prediction processes, enhancing accuracy by focusing on local data points and fuzzy clustering. In contrast, the second system relies on expert-provided assessments to shape the model, using analytic network processes and variable fuzzy sets to evaluate the severity of diabetes complications. This system integrates expert knowledge directly into the training phase, aligning with the core principle of RMLT by embedding domain-specific rules throughout the modeling process.

\subsubsection{Clinical Emphasis and Health Conditions}
The Hybrid algorithm of MCDM-VFS-IDM \cite{ahmadi2019novel} is tailored for analyzing the severity of damages caused by diabetes complications, integrating expert knowledge and multiple methodologies for comprehensive analysis. Similarly, FWNNet \cite{ahmadi2021fwnnet} focuses on the diagnosis of brain tumors using MRI images, combining fuzzy logic and neural networks to enhance both accuracy and explainability in complex tumor classification scenarios.

Additionally, the Fuzzy Gain Ratio Attribute Selection Method \cite{dai2013attribute} is designed for tumor classification based on gene expression data, employing fuzzy rough set theory and information gain ratio to effectively manage the challenges of high-dimensionality and small sample sizes. The Integrated Decision Support System based on ANN and Fuzzy AHP \cite{samuel2017integrated} predicts heart failure risks by integrating rule-based and machine learning approaches to assess the variable attributes contributing to heart failure, thereby improving prediction accuracy.

Furthermore, the Locally Weighted Factorization Machine with Fuzzy Partition \cite{zhou2022locally} predicts hospital readmission rates, especially for elderly patients, enhancing prediction accuracy through its adaptability to specific patient demographics. Lastly, a chart for the different categories can be seen in figure \ref{fig:hybrid_architectures}. The chart shows that within the RMLT dataset, Diagnostic Support Systems dominate, comprising three out of the five systems. In contrast, Decision Support Systems and Predictive Systems each account for one system, with no segmentation systems present.

\subsubsection{Strengths of RMLT Systems}
RMLT systems capitalize on the synergistic potential of rule-based reasoning and machine learning to enhance the precision of predictions and diagnoses significantly. For example, the Hybrid algorithm of MCDM-VFS-IDM \cite{ahmadi2019novel} integrates expert insights across its processes, making its outputs clinically relevant and practical. Similarly, FWNNet \cite{ahmadi2021fwnnet} achieves high accuracy and explainability by leveraging the strengths of fuzzy logic for handling uncertainty and neural networks for learning complex patterns. Moreover, systems like the Fuzzy Gain Ratio Attribute Selection Method \cite{dai2013attribute} successfully handle high-dimensional data and small sample sizes in gene expression, crucial for effective tumor classification. These systems also adeptly manage medical datasets fraught with uncertainty and incompleteness through methods like fuzzy logic, providing reliable predictions in challenging conditions. Furthermore, they excel in analyzing complex data interdependencies, essential for diseases like diabetes where multiple factors may influence the progression and treatment outcomes. Overall, the most common mentioned strength is the handling of complex data.

\subsubsection{Weaknesses of RMLT Systems}
Despite their robust capabilities, RMLT systems encounter significant operational challenges. The complexity of integrating diverse methodologies, such as those seen in the Hybrid algorithm of MCDM-VFS-IDM \cite{ahmadi2019novel} and the Fuzzy Gain Ratio Attribute Selection Method \cite{dai2013attribute}, increases system complexity, complicating maintenance and requiring specialized expertise. The example in Figure \ref{fig:ex2_rmlt} can be used again to show this. The creation and integration exists of many processes, as can be seen in the figure, further confirming this weakness. These systems' effectiveness heavily relies on the availability and quality of data, as demonstrated by the Hybrid algorithm of MCDM-VFS-IDM \cite{ahmadi2019novel}, where effectiveness is contingent on high-quality diabetes-related data. They also demand substantial computational resources, especially when processing complex datasets, which can be a limiting factor in resource-constrained environments. Additionally, the need for continuous updating and validation to keep pace with the rapid evolution of medical practices demands ongoing resource investment, as seen with systems like FWNNet \cite{ahmadi2021fwnnet}, which must integrate the latest advancements in neural network design and fuzzy logic.

\subsection{PERML}
\begin{figure}[htbp]
    \centering
    \includegraphics[width=0.9\textwidth]{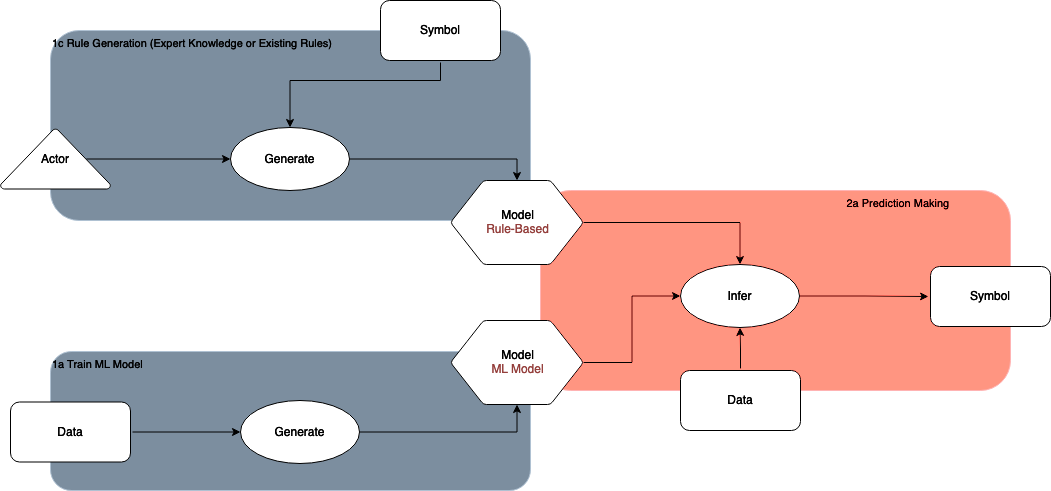}
    \caption{PERML}
    \label{fig:perml}
\end{figure}

\begin{figure}[htbp]
    \centering
    \includegraphics[width=0.9\textwidth]{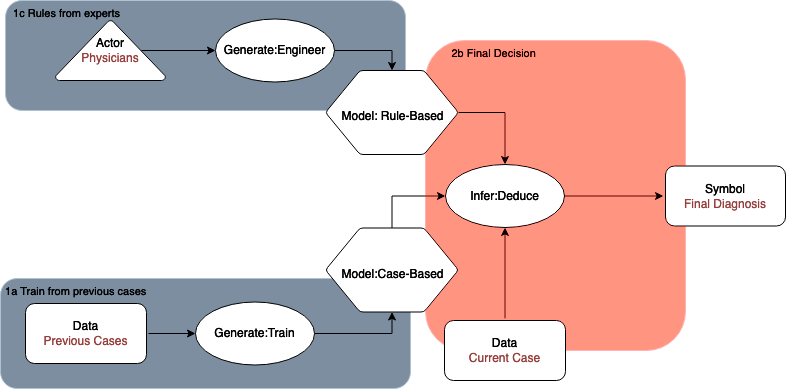}
    \caption{Knowledge Based Decision Support System for Detecting and Diagnosis of Acute Abdomen Using Hybrid Approach}
    \label{fig:ex1_perml}
\end{figure}

The primary motivation for the PERML system is to achieve high prediction accuracy. During the review, only one system was found that met the criteria: the Knowledge-Based Decision Support System (KBDSS) for diagnosing acute abdomen. A second system was not included because the paper only contained an abstract and lacked sufficient detail for evaluation. The generalized design pattern of PERML can be found in Figure \ref{fig:perml}. Figure \ref{fig:ex1_perml} shows the only system found. As can be seen, the system integrates both rule-based and case-based reasoning models. Data from previous cases and expert rules from physicians are used to train and engineer these models. The system infers and deduces the final diagnosis by combining inputs from both models, leveraging a hybrid reasoning approach to support physicians in making accurate diagnostic decisions.

\subsubsection{Clinical Emphasis and Health Conditions of PERML System}
The Knowledge-Based Decision Support System (KBDSS) for diagnosing acute abdomen conditions aims to enhance diagnostic accuracy. It focuses on the rapid and precise identification of acute abdominal causes, critical for emergency medical situations. The KBDSS combines rule-based and case-based reasoning. Rule-based reasoning applies predefined rules to diagnose conditions, while case-based reasoning compares new patient cases with previously diagnosed cases using cosine similarity metrics.  \cite{workneh2019knowledge}

\subsubsection{Strength of PERML System}
The KBDSS mentions several strong points. It achieves high diagnostic accuracy (99\% for cause classification, 66\% for severity identification), significantly outperforming physicians' rates (84\% and 50\%). The hybrid approach of combining rule-based and case-based reasoning boosts overall accuracy to 87.66\%. \cite{workneh2019knowledge}

The KBDSS enhances efficiency by providing quick, accurate diagnoses, crucial for urgent conditions like acute abdomen, facilitating timely treatment and potentially saving lives. It also supports better decision-making for physicians and encourages knowledge sharing among experts, improving training and diagnostic capabilities. \cite{workneh2019knowledge} 

\subsubsection{Weaknesses of PERML System}
However, the KBDSS has several weaknesses. Developing and integrating the system is complex, requiring expertise in rule-based and case-based reasoning, as well as proficiency in Prolog, Java, and MSQL. This complexity can hinder widespread adoption. Acquiring and encoding tacit knowledge from experts is challenging and time-consuming, which can slow progress. \cite{workneh2019knowledge}

Maintaining and updating the KBDSS to keep it current with medical knowledge is resource-intensive. The system’s performance heavily depends on the quality of data, in the case library. The KBDSS is tailored for acute abdomen, requiring significant modifications for other conditions, limiting its scope. \cite{workneh2019knowledge}

\subsection{New Patterns and Elementary Patterns}
\begin{figure}[htbp]
    \centering
    \begin{subfigure}[b]{0.45\textwidth}
        \centering
        \includegraphics[width=\textwidth]{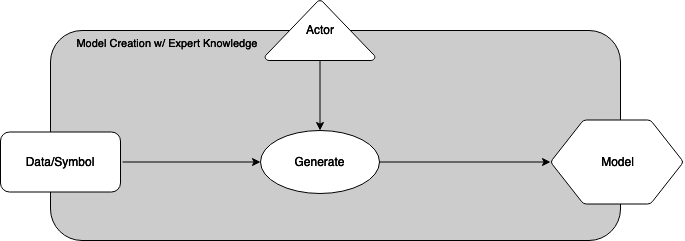}
        \caption{Expert Knowledge to Generate Model}
    \end{subfigure}
    \hfill
    \begin{subfigure}[b]{0.45\textwidth}
        \centering
        \includegraphics[width=\textwidth]{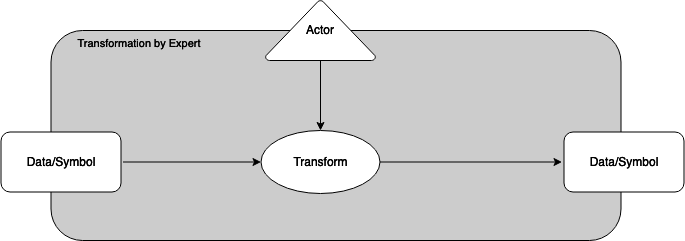}
        \caption{Expert Transforms Data}
    \end{subfigure}

    \vspace{0.5cm} 
    \begin{subfigure}[b]{0.45\textwidth}
        \centering
        \includegraphics[width=\textwidth]{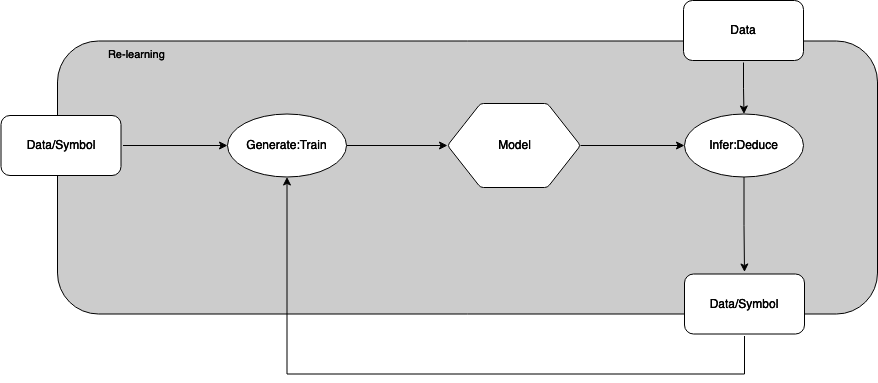}
        \caption{Re-learn with Outcome}
    \end{subfigure}
    \hfill
    \begin{subfigure}[b]{0.45\textwidth}
        \centering
        \includegraphics[width=\textwidth]{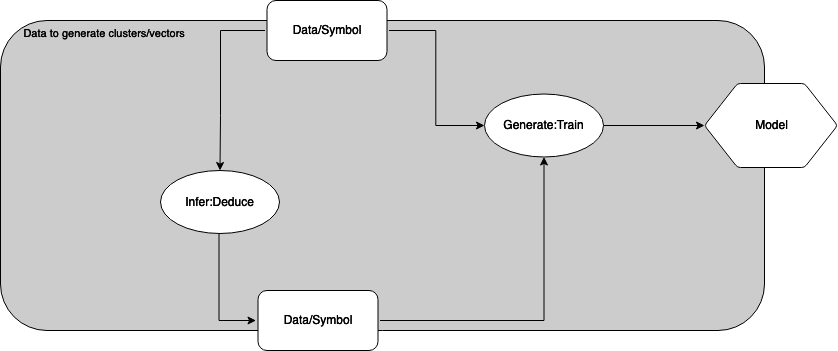}
        \caption{Generate Clusters and Vectors}
    \end{subfigure}

    \vspace{0.5cm} 

    \caption{Clinical Tasks}
    \label{fig:new_patterns}
\end{figure}

During the research, four new recurring patterns were identified that play a crucial role in the integration of expert knowledge with existing data in machine learning processes. These patterns can be seen in \ref{fig:new_patterns}

The first pattern involves the generation of models through the synergy of expert knowledge and existing data. In this approach, an expert uses their specialized knowledge in conjunction with available data to create new models. This combination ensures that the resulting models are of high quality and relevance, benefiting from the nuanced understanding that only human expertise can provide. 

The second pattern is the transformation of data by experts. This process includes tasks such as annotating or categorizing data points, which are critical for preparing data for subsequent processes. By incorporating expert input, the data is refined, making it more suitable for training machine learning models.

Re-learning represents another significant pattern where the output from a model is utilized as part of the training data for further iterations. This feedback loop allows for continuous improvement, as the model learns from its own predictions and adjusts accordingly. This iterative learning process is vital for maintaining the model's accuracy and relevance over time, ensuring that it evolves and adapts to new information.

The final pattern involves the use of data to generate clusters or vectors, which are then employed in combination with the original data to train a model. This method enhances the data's structure, providing additional context that improves the model's performance. By organizing the data into more meaningful units, this approach helps in uncovering hidden patterns and relationships within the data, leading to more robust and reliable models.

Table \ref{tab:freq} gives an overview of the occurrence of elementary patterns, mentioned in Bekkum et al., 2021. \cite{van2021modular}. Furthermore, the new patterns are also counted. The data shows that Pattern 1D is frequently used across all archetypes, with the highest occurrences in REML (36.7\%) and RMLT (51.9\%). Indicating that pre-processing or transforming data is important in those systems. Pattern 2A is also prominently featured, especially in MLRB (27.8\%) and RBML (26.2\%). Patterns such as 1B and 2C are less common, appearing infrequently across the archetypes. New patterns have notable representations in MLRB and RBML, indicating their relevance in these configurations. Lastly, there is no significant difference in the frequency of appearance for each pattern across the different archetypes.

\section{Discussion}
\subsection{Hybrid AI Systems}
Based on the strengths and weaknesses of different hybrid AI system architectures, specific recommendations can be made for their use in various healthcare tasks and conditions. REML systems excel in tasks where data is limited. Furthermore, REML has shown to be very adaptable with new data. These systems are mostly used for tasks involving segmentation. They are particularly effective in tasks such as, anomaly detection or image classification. This is reflected by the fact that the most commen health condition in these system is with regards to cancer, where data is often limited.

MLRB systems, on the other hand, are better for handling large datasets and complex data integration, with an emphasis on initial data preprocessing. These systems are particularly beneficial for early diagnosis of clinical events. By improving the relevance and accuracy of rule-based inferences through effective data preprocessing, MLRB systems are suitable for conditions involving extensive and varied data.

RBML systems prioritize explainability and trustworthiness, making them essential for diagnostic processes. They were used in cardiovascular disease risk assessment, breast cancer risk evaluation. These systems start with rule-based preprocessing followed by machine learning predictions, ensuring transparent and understandable decision-making processes. This is critical for building clinician trust, especially in diagnostic applications.

Furthermore, due to the boxology patterns made, it is found that MLRB and RBML systems had more steps for preprocessing when compared to other categories. This further indicates that these systems place a higher emphasis on the initial preparation and refinement of data, which is crucial for their specific strengths in handling complex datasets and providing explainable results.

RMLT systems are recommended for aligning with clinical knowledge and managing high-dimensional data. RMLT systems use rules to guide machine learning training, integrating expert knowledge into the learning process. This enables them to handle complex and high-dimensional datasets effectively, making them suitable for diseases with multifaceted data interactions. However, it should be noted that only five systems were found for RMLT, meaning that it might not be entirely generalizable. Additionally, the found RMLT systems typically incorporate more steps for processing data, indicating their ability to manage high complexity and enhance effectiveness in clinical settings.

Despite the limited data on PERML systems, the single instance analyzed indicates potential for high diagnostic accuracy and efficiency in urgent care, suggesting further exploration and development of parallel hybrid systems could be beneficial.

In conclusion, hybrid AI systems, when appropriately matched to the clinical task and context, can significantly enhance healthcare delivery, and facilitate more effective and trustworthy clinical decision-making processes.

\subsection{Boxology}
Boxology's structured approach and modular design patterns have been invaluable in this research on hybrid AI systems. By providing a comprehensive toolkit, Boxology has enabled the study to develop and analyze these systems with clarity. By abstracting specific technical details, Boxology has helped uncover hidden commonalities between different systems, allowing to identify shared strategies and techniques more effectively.

Furthermore, the research introduced four new patterns. These patterns contribute to the development and refinement of hybrid AI systems by offering novel approaches to integrating expert knowledge with machine learning processes. Additionally, five abstract patterns were created to categorize the diverse systems, enhancing the taxonomical organization of Boxology. The research also refined the generalized patterns to better align with the specific needs of clinical decision-making, ensuring that the resulting systems are more robust and applicable in real-world scenarios.

\subsection{Limitations}
One significant limitation of the research is the limited number of systems analyzed for the RMLT and PERML architectures. This constraint impedes the ability to generalize the findings and draw comprehensive conclusions about these architectures' effectiveness and applicability across various healthcare settings. Therefore, this study cannot adequately represent the diversity of potential implementations, limiting its overall impact and relevance.

Another notable limitation is the dependency on the existing framework provided by Kierner et al. (2023). \cite{kierner2023taxonomy} While this taxonomy offers a valuable foundation, it inherently restricts the exploration of potentially other hybrid AI system architectures that fall outside this predefined scope. By building exclusively upon this existing framework, the research might overlook other approaches in the field, potentially stifling the depth of the study.

\section{Conclusion}
In conclusion, this study provides a comprehensive analysis of hybrid AI systems' design patterns and their efficacy in clinical decision-making. By leveraging the framework of Boxology, the research systematically categorized and compared diverse hybrid architectures involving machine learning and rule-based reasoning, offering a high-level understanding of their structural foundations and practical applications in healthcare.

The research addressed two pivotal questions: how to categorize and map heterogeneous learning and reasoning systems against established design patterns, and how to extract insights through a comparative analysis of these systems. To categorize and map heterogeneous learning and reasoning systems, the study examined five primary architectures: REML, MLRB, RBML, RMLT, and PERML. Each architecture was analyzed for its distinct characteristics and categorized based on its structural and functional attributes. This approach allowed for a clear mapping of diverse hybrid AI systems to specific design patterns, demonstrating how different systems align with the Boxology framework.

For the comparative analysis, the study assessed the strengths and weaknesses of each architecture in various clinical tasks. REML systems were found to excel in segmentation tasks and handling datasets with limited data. MLRB systems' proficiency in handling large datasets and complex data integration made them ideal for early diagnosis and comprehensive data analysis. RBML systems emphasized explainability and trustworthiness, crucial for building clinician trust in diagnostic applications. RMLT systems effectively managed high-dimensional data and aligned closely with clinical knowledge, making them well-suited for complex, multifaceted medical conditions. Although only one PERML system was analyzed, it demonstrated potential for high diagnostic accuracy and efficiency in urgent care scenarios. These insights were derived by comparing the frequency and application of design patterns across different architectures.

This research introduced four new patterns, created five abstract patterns for categorizing systems, and refined existing patterns to better align with clinical decision-making needs. These contributions enhance the taxonomical organization of Boxology and provide novel approaches to integrating expert knowledge with machine learning processes. Boxology's structured approach and modular design patterns offer significant advantages in developing and analyzing hybrid AI systems. By providing a taxonomically organized vocabulary and abstracting specific technical details, Boxology reveals hidden commonalities, and promotes the creation of reusable solutions. These benefits enhance the performance and reliability of hybrid AI systems, making them more robust and easier to maintain, ultimately improving their application in clinical decision-making.

Despite the study's contributions, it faces limitations, particularly the limited number of systems analyzed for RMLT and PERML architectures and the dependency on an existing framework. Future research should expand the scope to include a broader range of hybrid AI system architectures, explore new approaches beyond the established taxonomy, and analyze more systems using Boxology.

Overall, this study underscores the critical role of hybrid AI systems in advancing healthcare delivery and highlights the potential of Boxology to drive further innovation in AI technology integration within the medical field. By continuing to refine these systems and methodologies, we can enhance clinical decision support, improve patient outcomes, and contribute to the ongoing evolution of AI-driven healthcare solutions.

%
%
\bibliographystyle{splncs04}
\bibliography{bibliography}

\newpage


\appendix
\section*{Appendix}

\section{Generalized Patterns}
The stages are color-coded for clarity: brown for preprocessing steps, pink for category-specific steps, blue for model creation, and green for output steps, including prediction and final post-processing.

\clearpage
\section{Clinical Tasks}

\clearpage
\section{New Patterns}

\clearpage
\section{Health Conditions}

\clearpage
\section{Pattern Frequency}
This is an overview of the frequency of elemantary patterns and new patterns found during the research. The meaning of the elemantary patterns can be found in Bekkum et al., 2021. \cite{van2021modular}

\begin{table}[h!]
\centering
\begin{tabular}{lccccc}
\toprule
 & REML (n = 33) & MLRB (n = 17)  & RBML (n = 12) & RMLT (n = 5) & PERML (n = 1)\\
\midrule
1A & 24 (14.5\%) & 12 (12.4\%) & 5 (8.2\%) & 4 (14.8\%) & 1 (33.3\%) \\
1B & 3 (1.8\%) & 2 (2.1\%) & 0 (0.0\%) & 0 (0.0\%) & 0 (0.0\%) \\
1C & 4 (2.4\%) & 2 (2.1\%) & 0 (0.0\%) & 2 (7.4\%) & 1 (33.3\%) \\
1D & 61 (36.7\%) & 33 (34.0\%) & 22 (36.1\%) & 14 (51.9\%) & 0 (0.0\%) \\
2A & 40 (24.1\%) & 27 (27.8\%) & 16 (26.2\%) & 3 (11.1\%) & 1 (33.3\%) \\
2B & 3 (1.8\%) & 4 (4.1\%) & 2 (3.3\%) & 1 (3.7\%) & 0 (0.0\%) \\
2C & 3 (1.8\%) & 2 (2.1\%) & 0 (0.0\%) & 0 (0.0\%) & 0 (0.0\%) \\
2D & 1 (0.6\%) & 1 (1.0\%) & 0 (0.0\%) & 1 (3.7\%) & 0 (0.0\%) \\
Ex. Model & 4 (2.4\%) & 6 (6.2\%) & 6 (9.8\%) & 0 (0.0\%) & 0 (0.0\%) \\
Ex. Data & 8 (4.8\%) & 2 (2.1\%) & 3 (4.9\%) & 0 (0.0\%) & 0 (0.0\%) \\
Re-learn & 6 (3.6\%) & 5 (5.2\%) & 1 (1.6\%) & 1 (3.7\%) & 0 (0.0\%) \\
Clusters/Rules & 9 (5.4\%) & 1 (1.0\%) & 6 (9.8\%) & 1 (3.7\%) & 0 (0.0\%) \\
\midrule
Total & 166 & 97 & 61 & 27 & 3 \\
\bottomrule
\end{tabular}
\caption{Frequency of Elementary Patterns}
\label{tab:freq}
\end{table}

\clearpage
\section{Refined Patterns}

The stages are color-coded for clarity: brown for preprocessing steps, pink for category-specific steps, blue for model creation, and green for output steps, including prediction and final post-processing.

\subsection{REML}

\clearpage
\subsection{MLRB}

\clearpage
\subsection{RBML}

\begin{figure}[htbp]
    \centering
    \includegraphics[width=0.9\textwidth]{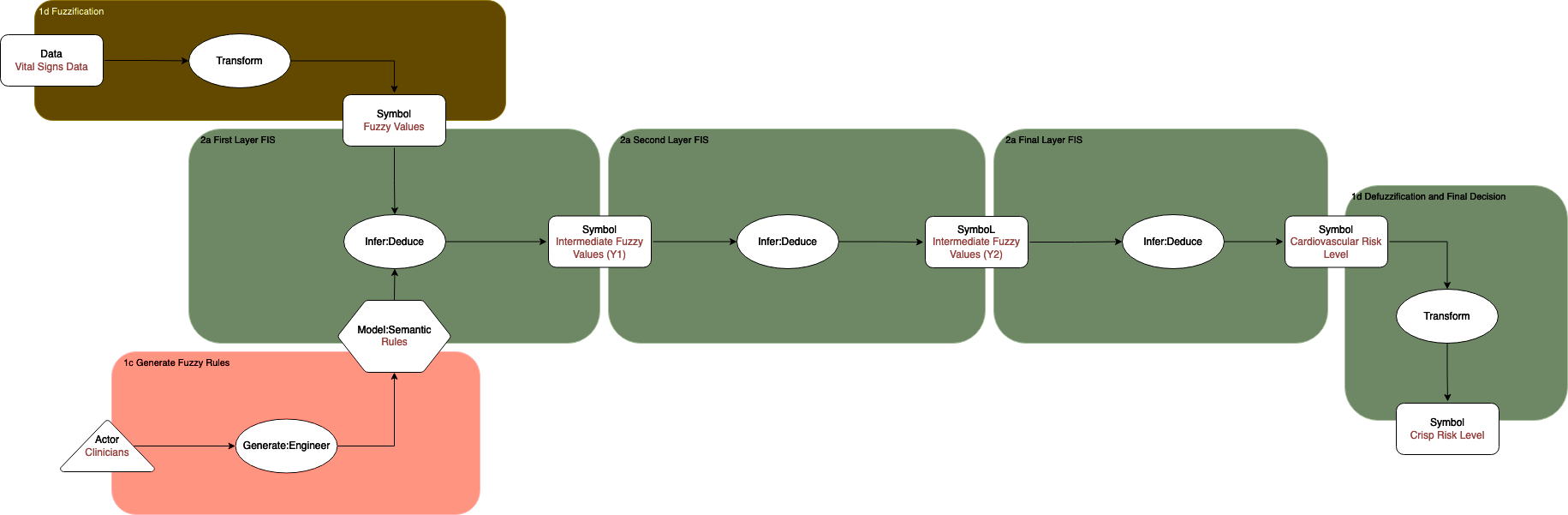}
    \caption{A Hierarchical Fuzzy System for Risk Assessment of Cardiovascular Disease}
    \label{fig:ex1_rbml}
\end{figure}

\begin{figure}[htbp]
    \centering
    \includegraphics[width=0.9\textwidth]{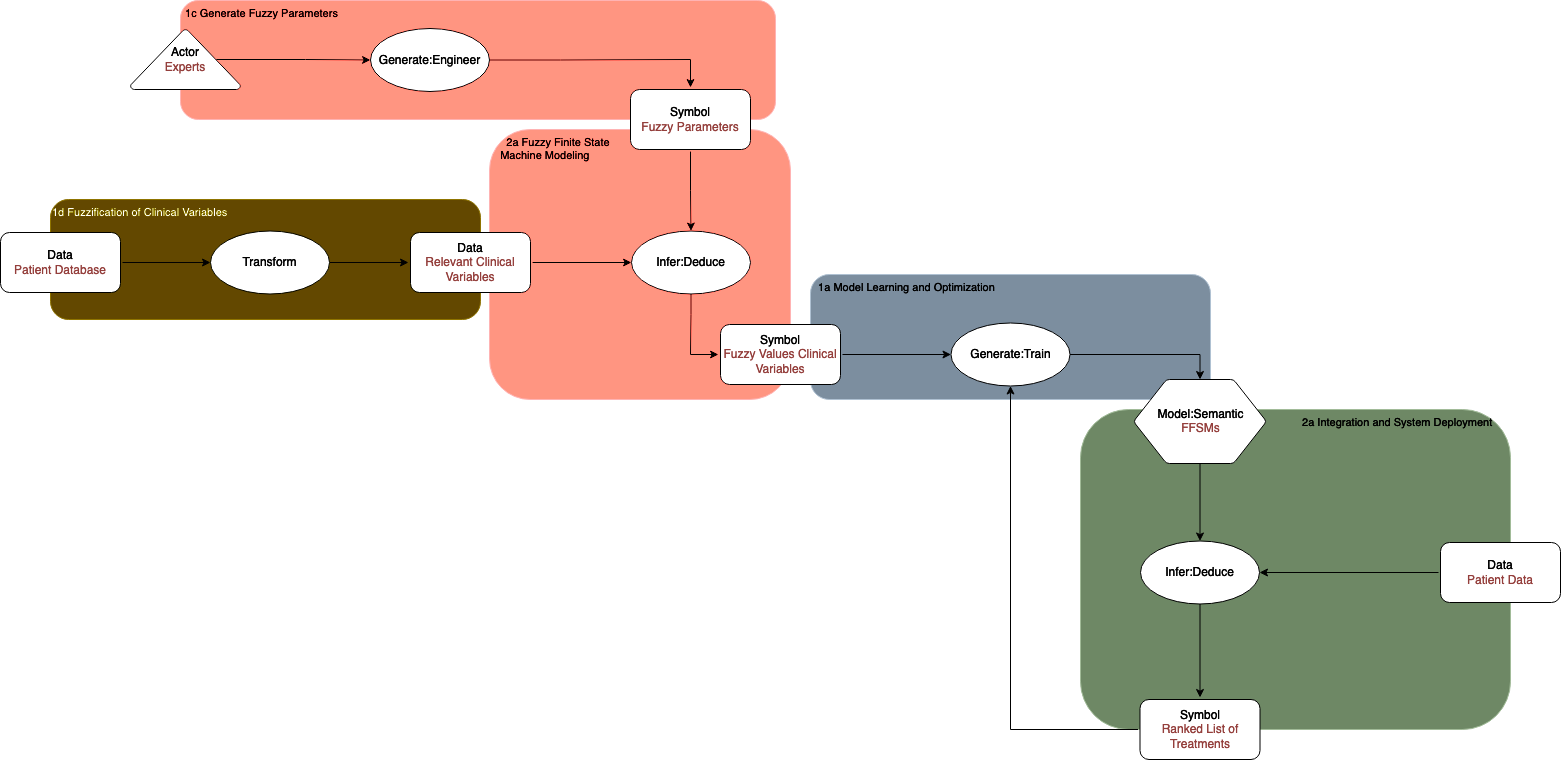}
    \caption{A Self-Learning Fuzzy Discrete Event System for HIV and AIDS Treatment Regimen Selection}
    \label{fig:ex4_rbml}
\end{figure}

\clearpage
\subsection{RMLT}

\clearpage
\subsection{PERML}


\end{document}